\documentclass[prb, reprint, amsfonts, amssymb, amsmath, superscriptaddress]{revtex4-2}
\usepackage[english]{babel}
\usepackage[utf8]{inputenc}
\usepackage{xcolor}
\usepackage{graphicx}
\usepackage{multirow}
\usepackage{xspace}
\usepackage{times}
\usepackage{hyperref}
\hypersetup{colorlinks=true, linkcolor=black, citecolor=green!50!black, urlcolor=blue!50!black}

\let\hatOrig\hat
\renewcommand{\vec}[1]{\boldsymbol{\mathbf{#1}}}
\renewcommand{\hat}[1]{\boldsymbol{\mathbf{\hatOrig{#1}}}}

\newcommand{\sub}[1]{\ensuremath{_{\textrm{#1}}}} 
\newcommand{\super}[1]{\ensuremath{^{\textrm{#1}}}} 

\newcommand{\affilMSE}{\affiliation{Department of Materials Science \& Engineering, Rensselaer  Polytechnic Institute, 110 8$^{th}$ St, Troy, NY 12180}}
\newcommand{\affilIBMAlbany}{\affiliation{IBM Research, 257 Fuller Road, Albany, NY 12203}}
\newcommand{\affilIBMtext}{IBM Thomas J. Watson Research Center, 1101 Kitchawan Road, Yorktown Heights, NY 10598}
\newcommand{\affilIBM}{\affiliation{\affilIBMtext}}

\newcommand{\affilNCKU}{\affiliation{National Cheng Kung University, Tainan City, Taiwan}}
\newcommand{\affilAS}{\affiliation{Institute of Physics, Academia Sinica, Taipei, Taiwan}}
\newcommand{\affilNUS}{\affiliation{Electrical and Computer Engineering, National University of Singapore, Singapore}}
\usepackage{lineno}

\begin{document}

\title{Surface-dominated conductance scaling in Weyl semimetal NbAs}

\author{Sushant Kumar}
\affilMSE \affilIBM

\email{Sushant.Kumar@ibm.com, cchen3@us.ibm.com}

\author{Yi-Hsin Tu}
\affilNCKU

\author{Luo Sheng}
\affilNUS

\author{Nicholas A. Lanzillo}
\affilIBMAlbany

\author{Tay-Rong Chang}
\affilNCKU

\author{Gengchiau Liang}
\affilNUS

\author{Ravishankar Sundararaman}
\affilMSE

\author{Hsin Lin}
\affilAS

\author{Ching-Tzu Chen}
\affilIBM

\begin{abstract}
Protected surface states arising from non-trivial bandstructure topology in semimetals can potentially enable advanced device functionalities in compute, memory, interconnect, sensing, and communication. This necessitates a fundamental understanding of surface-state transport in nanoscale topological semimetals. Here, we investigate quantum transport in a prototypical topological semimetal NbAs to evaluate the potential of this class of materials for beyond-Cu interconnects in highly-scaled integrated circuits. Using density functional theory (DFT) coupled with non-equilibrium Green’s function (NEGF) calculations, we show that the resistance-area $RA$ product in NbAs films decreases with decreasing thickness at the nanometer scale, in contrast to a nearly constant $RA$ product in ideal Cu films.
This anomalous scaling originates from the disproportionately large number of surface conduction states which dominate the ballistic conductance by up to 70$\%$ in NbAs thin films.
We also show that this favorable $RA$ scaling persists even in the presence of surface defects, in contrast to $RA$ sharply increasing with reducing thickness for films of conventional metals, such as Cu, in the presence of surface defects.
These results underscore the potential of topological semimetals as future back-end-of-line (BEOL) interconnect metals. 
\end{abstract}

\maketitle



\section{Introduction}
\begin{figure*}[t!]
\centering
\includegraphics[width=\textwidth]{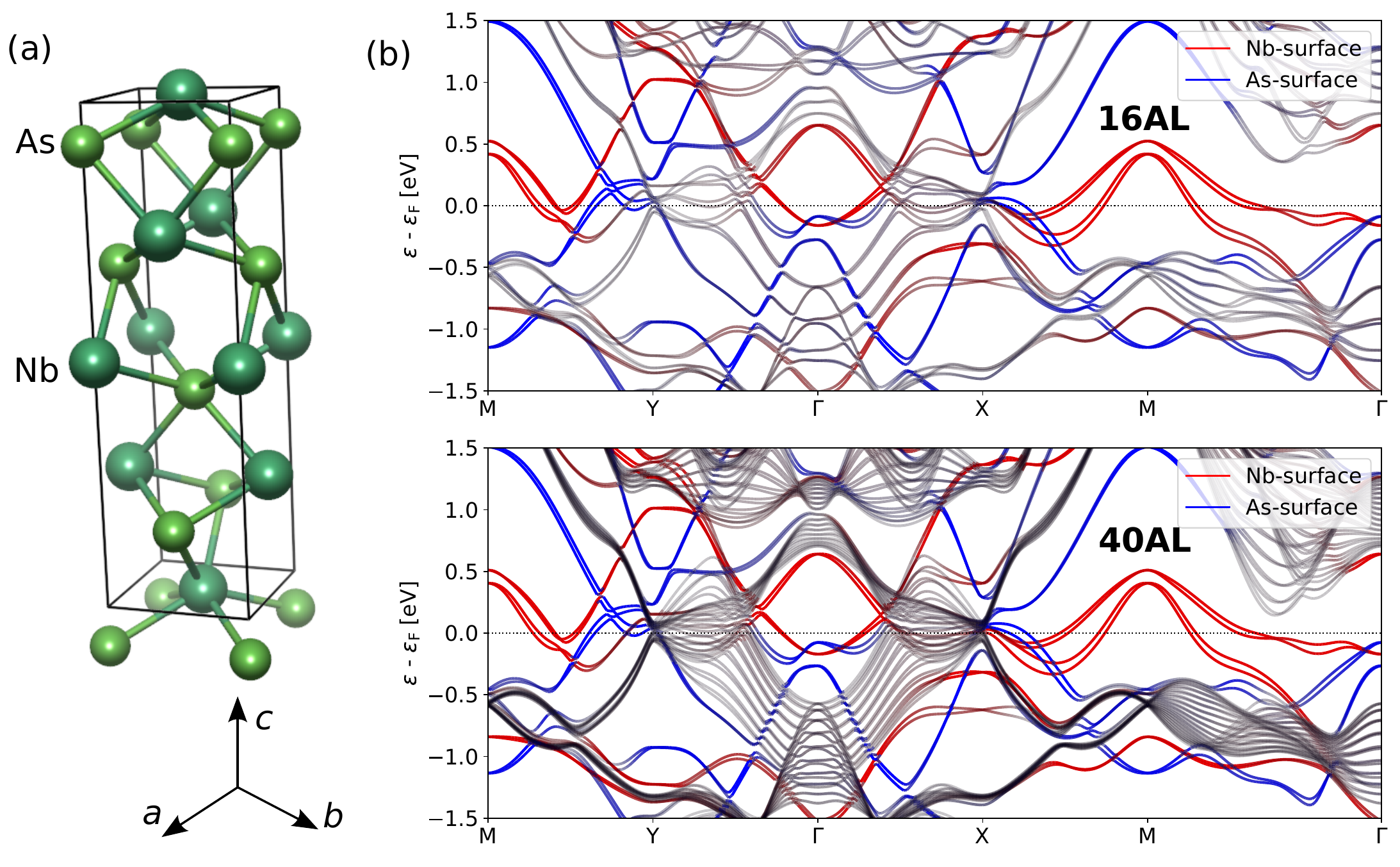}
\caption{\textbf{Crystal structure of bulk NbAs and bandstructures of thin films of NbAs} (a) The tetragonal unit cell of NbAs comprises 8 atomic layers such that each Nb and As atom has a coordination number of 6. The crystal has time-reversal symmetry but lacks a space-inversion symmetry. When a bulk NbAs crystal is cleaved along (001) surface, it produces a Nb-terminated surface (top) and an As-terminated surface (bottom). (b) DFT bandstructures for 16AL and 40AL (001) films of NbAs with colors representing the contribution of the bulk (gray), Nb-terminated (red) and As-terminated (blue) surfaces to the electronic states. Increasing the thickness of slabs, increases the number of bulk bands at the Fermi level though the surface bands remain largely unchanged.   }
\label{fig:intro} 
\end{figure*}
The discovery of Weyl semimetals and topological semimetals in general~\cite{xu2015discovery,yang2015weyl,lv2015experimental,xu2015discoveryNbAs,armitage2018weyl,lv2021experimental,hasan2021weyl,yan2017topological,bansil2016colloquium} has prompted research into 
the discovery of new phenomena and their applications
in various areas of condensed matter physics.
These include usage as 
far-infrared and tetrahertz detectors~\cite{wu2017giant}, magnetoresistive memory devices~\cite{han2021current,de2021gigantic}, photovoltaic devices~\cite{osterhoudt2019colossal}, and as interconnects in next-generation integrated circuits (ICs)~\cite{chen2020topological,han20211d,gall2021materials,lanzillo2022size}.
A Weyl semimetal can be formed by breaking either time-reversal or inversion symmetry in a crystal with 3D Dirac cones, leading to pairs of band crossing points called Weyl nodes.
The surface Brillouin Zone of Weyl semimetals have projections of such Weyl node pairs connected through series of topologically-protected Fermi-arc surface states~\cite{lv2015experimental}.
Substantial recent research efforts have targeted first-principles prediction of new topological semimetals, material syntheses, and confirmation of nontrivial band structures and Fermi-arc surface states using angle-resolved photoemission spectroscopy (ARPES)~\cite{bansil2016colloquium,armitage2018weyl,lv2021experimental,hasan2021weyl,yan2017topological}.
These materials have been shown to exhibit unconventional transport, optical and magnetic phenomena~\cite{nagaosa2020transport,hu2019review,wang2017quantum,gorbar2018anomalous}, including chiral anomalies~\cite{ong2021experimental}, a nonlinear Hall effect~\cite{sodemann2015quantum,ma2019observation,kang2019nonlinear}, a quantized circular photogalvanic effect~\cite{de2017quantized,rees2020helicity,ni2021giant} and giant second-harmonic generation~\cite{wu2017giant,patankar2018resonance}.

Like topological insulators, the surface states of topological semimetals have received considerable attention, which if topologically protected, could potentially lead to high surface conduction.
Previous theoretical work has argued that the Fermi-arc states in a toy-model Weyl semimetal contribute the same order of magnitude as the bulk states to total conduction~\cite{breitkreiz2019large} and could be highly disorder tolerant when the Fermi arcs are nearly straight~\cite{resta2018high}.
However, other studies have shown that the transport due to Fermi arcs is dissipative due to a strong hybridization of surface and bulk states, which leads to scattering between surface and bulk states~\cite{gorbar2016origin,wilson2018surface}.
Since these studies relied primarily on highly-simplified Hamiltonians and analytical models, a comprehensive study of transport fully accounting for the electronic structure at dimensions relevant to future device applications is now necessary.


In this work, we pursue a fundamental understanding of electron transport properties of Weyl semimetals at nanoscale and evaluate their potential as high conductivity future interconnect metals.
In modern-day ICs, the devices patterned on a silicon substrate are linked to form a circuit using Cu nanowires called interconnects.
The resistivity of Cu increases dramatically with decreasing size due to enhanced scattering of electrons from surfaces, defects, and grain boundaries~\cite{fuchs1938math,sondheimer1952adv,mayadas1969electrical,mayadas1970electrical}.
Such increase in the resistivity can increase the signal delay and energy consumption by $\sim 40\times$, a major bottleneck in the semiconductor industry~\cite{gall2021materials,gall2020search}.
The search to replace Cu has expanded from elemental metals to  intermetallics~\cite{soulie2021aluminide,chen2018nial,chen2021interdiffusion,zhang2022resistivity}, metallic carbides and nitrides such as MAX phases~\cite{sankaran2021ab,zhang2021resistivity}, directional conductors~\cite{DirectionalConductors}, and topological materials~\cite{chen2020topological,han20211d,han2023topological,lien2023unconventional,lanzillo2022size}. 

In a recent breakthrough, Zhang et al. \cite{zhang2019ultrahigh} showed experimentally that the electrical resistivity of nanobelts of [001] oriented NbAs, a Weyl semimetal, becomes an order of magnitude lower than the bulk single-crystal resistivity.
In some nanobelt samples, the resistivity can even be lower than the bulk resistivity of Cu.
Such an anomalous reduction was attributed to transport via the disorder-tolerant Fermi-arc surface states in NbAs. %
Furthermore, using first-principles calculations, Chen et al. ~\cite{chen2020topological} predicted that thin films of a prototypical chiral topological semimetal CoSi can exhibit conduction dominated by Fermi-arc surface states, leading to an resistance-area ($RA$) product that decreases with decreasing thickness, in stark contrast to Cu and other conventional metal films.

Despite the promising trend of decreasing $RA$ product with dimensions, CoSi is still at a disadvantage compared to Cu because of the low density of states at the Fermi level and a significantly higher bulk resistivity. 
Hence, we need semimetals with larger numbers of topologically-protected surface states~\cite{chen2020topological,lanzillo2022size}.
The aforementioned Weyl semimetal NbAs is one such candidate with 12 pairs of Weyl nodes.
In this work, we use first-principles quantum transport calculations to predict the $RA$ product scaling of (001) NbAs thin films with and without surface defects.
We find that the $RA$ product decreases with decreasing film thickness for both pristine and defect-laden films, as previously shown for CoSi~\cite{chen2020topological}.
However, NbAs does not exhibit the protection of surface transport protected against line-defects that was shown for CoSi due to the chiral nature of its surface states.
Our calculations illustrate that the observed $RA$ scaling in NbAs is due to the large number of surface states that account for at least $50\%$ of conduction for films thinner than $\sim7$ nm.
The contribution of the Nb-terminated surfaces in (001) NbAs films is roughly 3 times that of the As-terminated surfaces.
Lastly, we show that surface-mediated conduction and favorable $RA$ scaling with thickness survives in the presence of minor surface disorder.

\begin{figure*}[t]
\centering
\includegraphics[width=\textwidth]{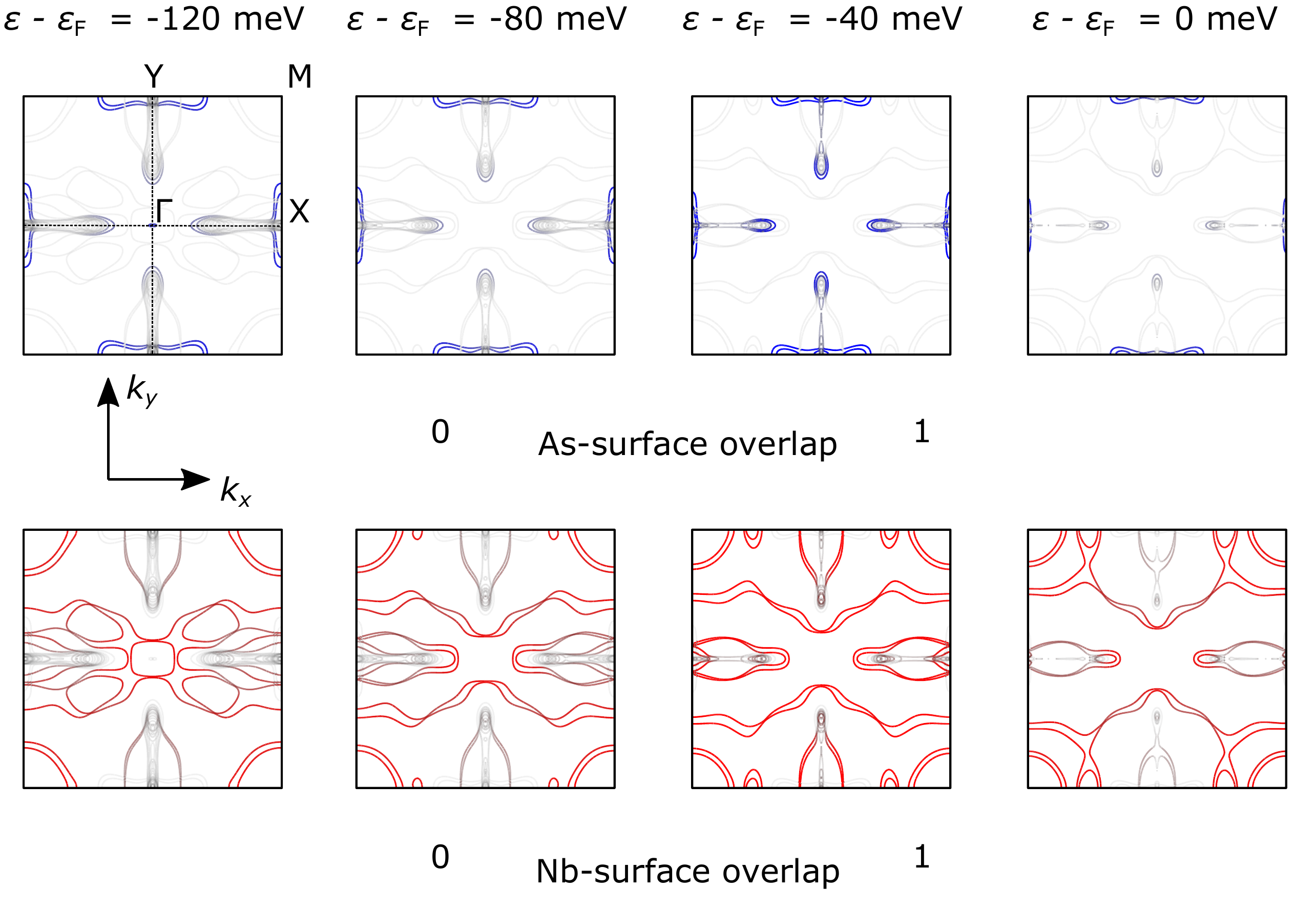}
\caption{\textbf{Fermi surfaces of NbAs slabs} Isoenergy surfaces for a relaxed 56AL (001) slab of NbAs (thickness $\sim 80.24$  \AA) at energies $\varepsilon = 120, 80, 40$ and $0$~meV below the neutral Fermi level $\varepsilon_\mathrm{F}$.
The colors represent the contribution of the bulk (gray), Nb-terminated surface (red) and As-terminated surface (blue) to the electronic states.
The Nb-terminated surface contributes many more states that extend throughout the Brillouin zone, compared to the fewer As-terminated surface states that are localized in the $\mathbf{k}$-space. 
The Fermi arc states agree with ARPES measurements best for $\varepsilon - \varepsilon_\mathrm{F} = -80$~meV.}
\label{fig:FS}
\end{figure*}

\section{Results}

\textbf{Bandstructure and Fermi surface:} Figure~\ref{fig:intro}(b) shows the first-principles-computed bandstructures of 16 atomic-layer (AL) ($\sim 21.43 \textup{~\AA}$) and 40 AL ($\sim 56.71 \textup{~\AA}$) (001) slabs of NbAs. The colors represent the contribution of spatial regions to each electronic state: bulk in gray, Nb-terminated surface in red and As-terminated surface in blue. Increasing the thickness of the slabs (16 AL $\rightarrow$ 40 AL) increases the number of bulk bands but the surface bands remain largely unchanged. Note that the (001) surface of NbAs reduces the $C_{4}$ rotational symmetry of the bulk to $C_{2}$~\cite{xu2015discovery,sun2015topological}. As a result, both the Nb-terminated (red) and As-terminated (blue) surface bands differ between the $\Gamma$-X and $\Gamma$-Y high-symmetry $\mathbf{k}$-point paths. The bulk (gray) bands, which dominate the Y-$\Gamma$-X path, however, are mostly symmetric about $\Gamma$. At the Fermi level, the Nb-terminated surface bands are hole-like along X-$\Gamma$-Y and electron-like along M-Y-$\Gamma$-X. These results agree with previous DFT bandstructure calculations for NbAs films~\cite{sun2015topological}.

Next, we analyze the Fermi surfaces of (001) NbAs slabs to get insight into its electronic bandstructure.
Since electronic states at the Fermi level dominate conduction, we aim to find the chemical potential at which the DFT-predicted isoenergy surfaces agree the best with ARPES data~\cite{xu2015discoveryNbAs} to use for subsequent non-equilibrium Green's function (NEGF) calculations.
Figure~\ref{fig:FS} shows the isoenergy surfaces for a 56 AL slab computed using the Wannierized electronic states ($\mathbf{k}$-point grid: $512\times512)$ at different energy levels $\varepsilon$ near the neutral Fermi level $\varepsilon_\mathrm{F}$ , with  $\varepsilon - \varepsilon_\mathrm{F} \in \{-120, -80, -40, 0\}$~meV.
As described in the Methods section, these isoenergy surfaces have been resolved by contributions of the bulk (gray), Nb-terminated (red) and As-terminated (blue) surfaces. 

Comparing the top and bottom rows, we find that a disproportionate number of states belong to the Nb-terminated surface.
Sun et al.~\cite{sun2015topological} noted that strong hybridization between the surface and bulk states, and between trivial Fermi surfaces and arcs, makes it difficult to isolate the topological Fermi-arc states.
However, careful analysis of spin textures~\cite{sun2015topological} and ARPES measurements~\cite{xu2015discoveryNbAs} has indicated that the outer arc of the spoon-shaped features along $\Gamma$-X and $\Gamma$-Y are the Fermi arcs.
These arcs are not clearly visible at the DFT-predicted Fermi level $\varepsilon_\mathrm{F}$, but become clearer with decreasing $\varepsilon$ and achieve good agreement with the ARPES measurements of the As-terminated surface at $\varepsilon - \varepsilon_\mathrm{F} = -80$~meV. Hence, we shift the Fermi level down by 80 meV from the original DFT-computed value for all transport predictions reported below. 

\begin{figure}[t!]
\includegraphics[width=\columnwidth]{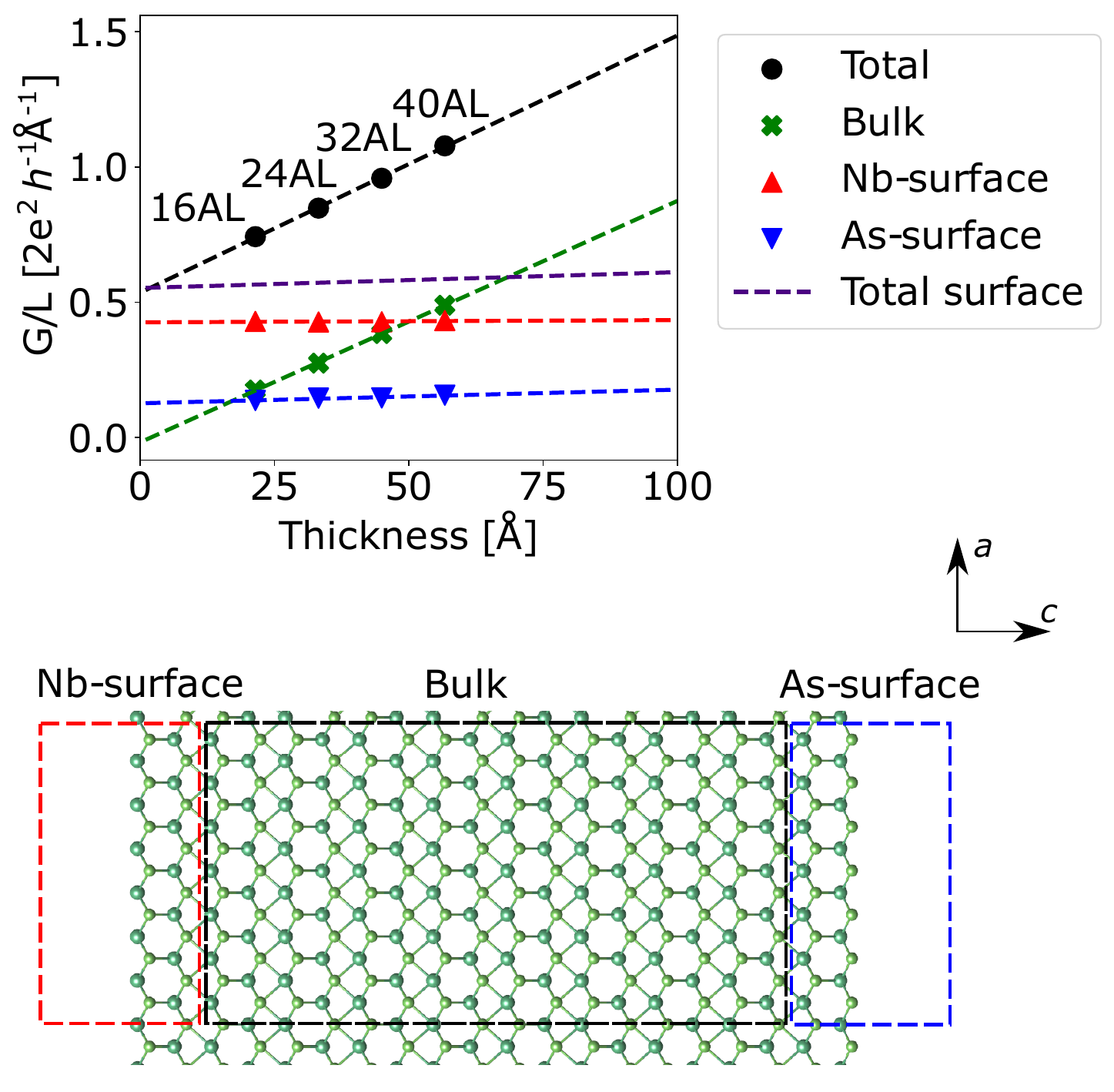}
\caption{\textbf{Contribution of the bulk and surface to the net ballistic conductance} Ballistic conductance as a function of thickness for pristine (001) NbAs films.
The total conductance increases linearly due to the linear increase in bulk conductance, while the conductance contribution due to surface states remains constant.
The dashed lines are linear fits to the computed data points for 16, 24, 32 and 40 AL slabs.
Surface conduction dominates over bulk conduction for slabs thinner than 6.8 nm, reaching $76\%$ of the total conduction for 2-nm-thick slabs.}
\label{fig:conductance_contrib} 
\end{figure}

\textbf{Ballistic conductance scaling:} We first compute the ballistic conductance of pristine films of NbAs as a function of thickness using
\begin{align}
    G = \frac{\mathrm{e}^2}{2}\int_{\mathrm{BZ}} \sum_b g_s\frac{d\vec{k}}{(2\mathrm{\pi})^3} f'_0(\varepsilon_{\vec{k}b})|v^x_{\vec{k}b}|
    \label{eqn:ballistic}
\end{align}
where $\varepsilon_{\vec{k}b}$ and $\vec{v}_{\vec{k}b}$ are the electronic energies and velocities of band $b$ and wavevector $\vec{k}$ in the Brillouin zone (BZ) and $g_s$ is the spin degeneracy factor ($g_s=1$ for the $\varepsilon_{\vec{k}b}$ vs. $\mathbf{k}$ relations with spin-orbit coupling). The derivative of the Fermi-Dirac occupations $f'_0(\varepsilon_{\vec{k}b})$ limits the contributions of electronic states to within a few $k_BT$ around the Fermi level $\varepsilon_\mathrm{F}$ (Refer SI for a detailed derivation of Equation~\ref{eqn:ballistic}).
We evaluate the above expression in JDFTx~\cite{JDFTx} for room temperature ($k_BT=0.026$ eV) using a Monte Carlo sampling of 250,000 $\mathbf{k}$-points in the BZ.
We also decompose the total conductance $G$ into contributions from bulk, Nb-terminated and As-terminated surfaces by weighing each electronic state in the integrand of Eq.~\ref{eqn:ballistic} with a slab weight function described in the Methods sections. The bounding box or slab used to define the spatial region for the surface states has been shown in Figure~\ref{fig:conductance_contrib}. Note that we use this approach to predict ballistic conductance in pristine films, which would have no scattering.

We see that the total ballistic conductance per unit length ($G/L$) increases linearly with thickness (Figure~\ref{fig:conductance_contrib}).
The decomposition of the total conductance into bulk and surface contributions shows that the Nb- and As-terminated surface-state contributions remain constant with thickness. This thickness independence of the surface states (and hence, the conductance) is supported by the electronic bandstructures of the slabs (Figure 1(b)) and the Fermi-surface plots resolved by the contribution of each atomic layer (Supplementary Figures 12 and 13). As noted in the previous section, the surface states (red and blue bands) remain remarkably the same as we increase the thickness from 16AL to 40AL. We only see an increase in the number bulk bands (gray). Supplementary Figures 12 and 13 show that the penetration depths of the surface states are roughly 0.6 nm ($\sim$ 6AL) for both the Nb- and As-terminated surfaces. Hence, the surface states would remain untangled as long as the thickness of the slabs is more than 12AL (i.e., slabs with 2 or more unit cells along the $z$-direction). Consequently, for slabs of such thickness there would be almost no hybridization of the states emanating from the opposite surfaces.

The bulk conductance contribution decreases linearly with decrease in film thickness and extrapolates to nearly zero for zero thickness. Hence, the total conductance $G$ can be expressed as
\begin{align}
G = g\sub{bulk} t
+ G\super{Nb}\sub{surf}+G\super{As}\sub{surf}
\label{eqn:g_fit}
\end{align}
where $t$ is the thickness of the film, $g\sub{bulk}$ is the slope of the linear fit to bulk conductance and $G\super{Nb}\sub{surf}$ and $G\super{As}\sub{surf}$ are the conductance due to Nb- and As-terminated surfaces respectively. For a 16 AL ($\sim$ 2.1 nm) slab, the surface states and bulk account for $76.3\%$ and $23.7\%$ of the total ballistic conductance respectively. 
Such large surface state contributions to conductance have been observed for other topological semimetals as well, e.g. $\sim 90~\%$ surface-state contribution in 2.7-nm-thick CoSi~\cite{lien2023unconventional}.
As we increase the thickness to 40 AL ($\sim$ 5.7 nm), the bulk conductance contribution for NbAs increases to $45.4\%$ while the surface contribution reduces to $54.6\%$.
Extrapolation of the linear fits to bulk and total surface conductance ($G\super{Nb}\sub{surf}+G\super{As}\sub{surf}$) reveals that the crossover point where surface and bulk conductance become equal is at around 6.8 nm which corresponds to a relaxed 48 AL slab.
We also find that due to the larger number of states at the Fermi level, the ballistic conductance of NbAs (001) films is larger than that of CoSi (See Supplementary Figure 8).
Specifically, for a 2.5-nm-thick slab, the conductance for NbAs is around $57\%$ higher than that of CoSi.    

Importantly, the Nb-terminated surface contributes almost 3 times as much as the As-terminated surface to ballistic conductance, i.e., $G\super{Nb}\sub{surf} \approx 3G\super{As}\sub{surf}$.
This is in line with the Nb-terminated surface states vastly outnumbering the As-terminated surface states in the surface-resolved Fermi surfaces shown in Figure~\ref{fig:FS}. 

Note that shifting the boundary of the bounding box/slab further into the slab (Figure~\ref{fig:conductance_contrib}) would count more electronic states as surface states, including a part of the bulk conductance into the surface.
While the definition of these boundaries is arbitrary, we have chosen  it to the maximum value for which $G^{\mathrm{Nb/As}}\sub{surf}$ remains thickness independent in order to capture as much of the surface state contribution as possible, without including the bulk.

\begin{figure}[t!]
\includegraphics[width=\columnwidth]{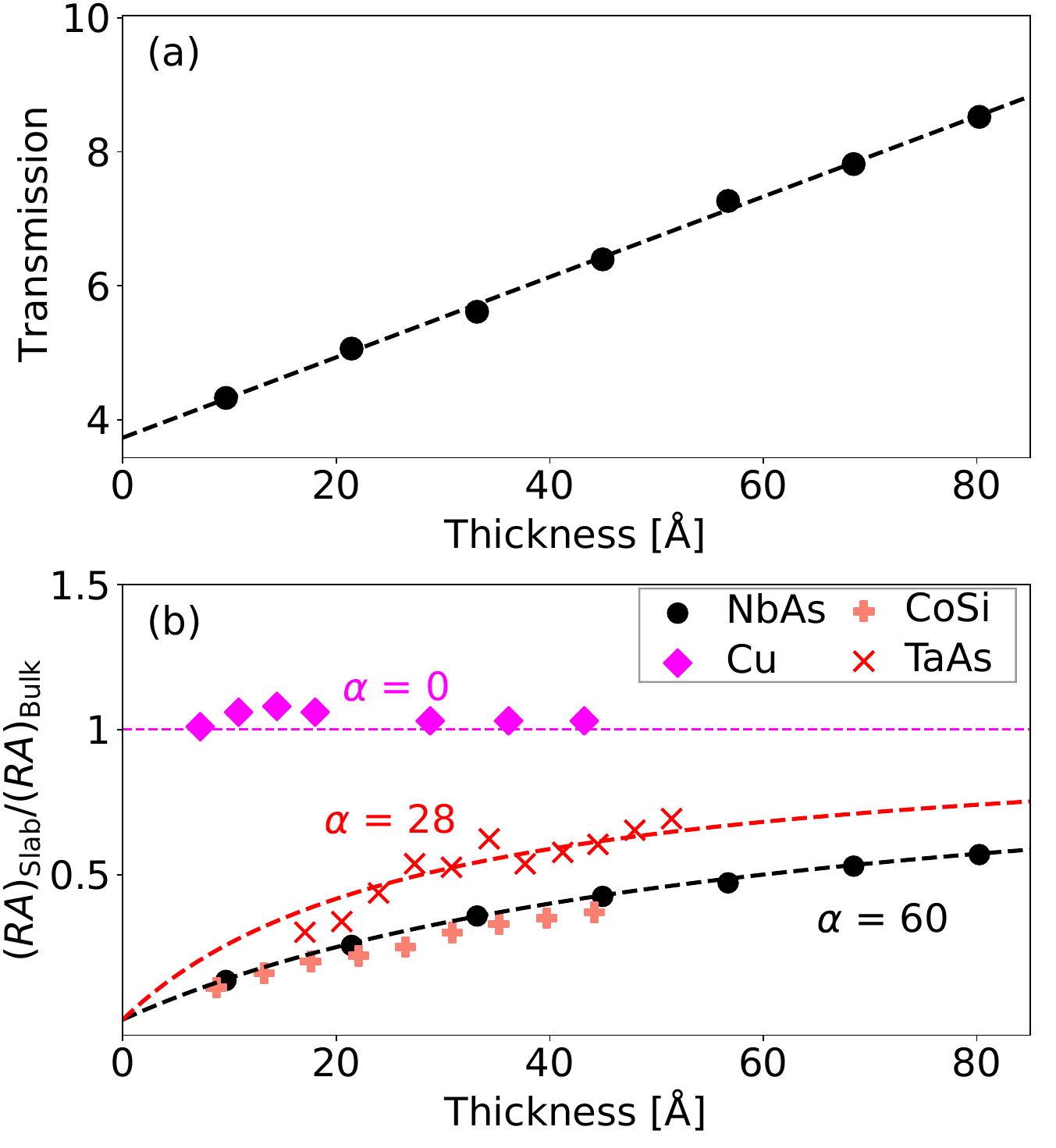}
\caption{\textbf{Comparison of the ballistic conductance of pristine thin films of different materials} (a) Transmission as a function of thickness for pristine (001) slabs of NbAs. The non-zero intercept corresponds to surface conduction.
(b) Normalized resistance-area $RA$ product for transport along [100] direction of NbAs (001) slabs, compared against [001] transport direction of (100)-terminated surfaces of CoSi \& Cu and [001] transport direction for (100)-terminated surface of TaAs taken from Ref.~\citenum{chen2020topological}.
The dashed line is the fit of Equation~\ref{eqn:RA_fit} to the computed conductance data for NbAs. 
Topological semimetals NbAs and CoSi show a promising trend of decreasing $RA$ product with decreasing thickness due to the significant contribution of surface states to total conductance for thin films of both materials. }
\label{fig:RA} 
\end{figure}

\textbf{Resistance-area product scaling:} We next analyze the resistance-area ($RA$) product scaling for films of NbAs with and without defects (pristine) and compare the results with those of Cu (a conventional metal) and CoSi (a chiral multifermion semimetal) (Figure~\ref{fig:RA}(a)).
The resistance $R$ of these films have been calculated using the transmission $T$ at the Fermi level $\varepsilon_\mathrm{F}$ using
\begin{align}
    R = \frac{1}{G_0 T(\varepsilon=\varepsilon_\mathrm{F})}
\end{align}
Here, $G_0$ is the quantum of conductance $\mathrm{e}^2/h$. The transmission is computed using the NEGF method~\cite{datta2005quantum}, where we employ Wannier tight-binding Hamiltonians constructed using DFT as described in the Methods section. 

Previous first-principles NEGF calculations have shown that the $RA$ product of slabs $(RA)\sub{slab}$ for pristine Cu is mostly independent of slab thickness~\cite{timoshevskii2008influence, chen2020topological,lien2023unconventional}, because bulk states dominate conduction.
A similar trend has also been observed for MoP, a topological metal, where most of the electronic states at the Fermi level are bulk states~\cite{han2023topological}.
Hence, for such materials, conductance $G(=1/R)$ decreases linearly with decreasing thickness or cross-sectional area $A$, making the $RA$ product constant.
Consequently, independent of film thickness, the normalized $RA$ product $(RA)\sub{slab}/(RA)\sub{bulk} \approx 1$.

\begin{figure*}[t!]
\centering
\includegraphics[width=1.\textwidth]{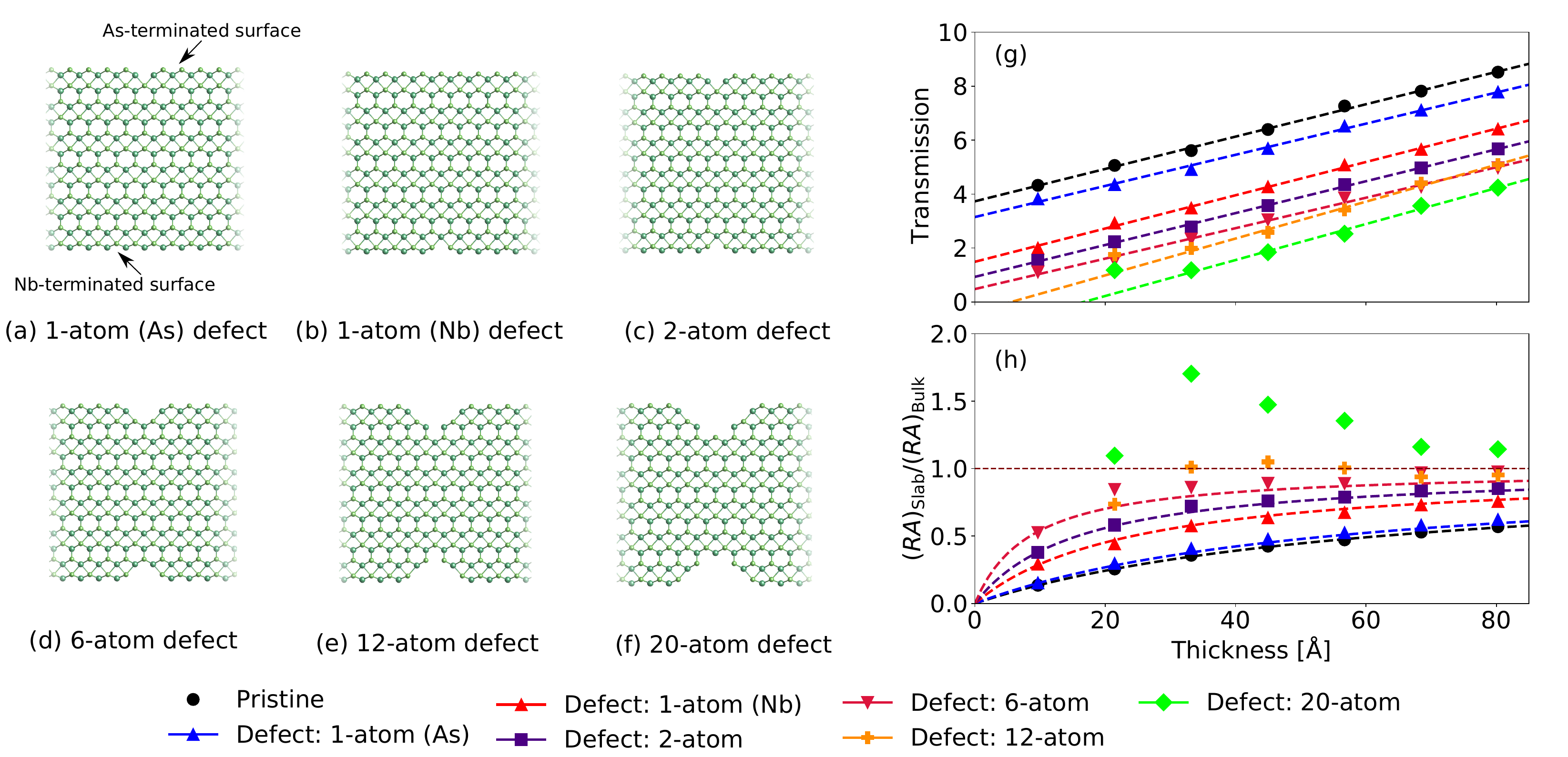}
\caption{\textbf{Effect of different defect configurations on the electron transport in NbAs slabs} (a-f) Illustration of the different surface line-defect configurations studied in our NEGF calculations.
(g) Transmission and (h) normalized resistance-area $RA$ product for [100] transport in NbAs(001) with defects. 
Line defects reduce the net transmission, but the intercept remains nonzero and the corresponding $RA$ remains below the bulk value, indicating that surface conduction persists.
Only when the notch is made deep enough, such as for the 20-atom defect, the $RA$ product begins to increase with decreasing thickness due to a significant reduction in the transmission.}
\label{fig:defects} 
\end{figure*}

In contrast, NEGF calculations of pristine films of NbAs show that $(RA)\sub{slab}$ decreases with decreasing film thickness and is always less than $(RA)\sub{bulk}$, similar to previous reports for CoSi films~\cite{chen2020topological, lien2023unconventional}.
This can be explained by extending Equation~\ref{eqn:g_fit} to calculate $(RA)\sub{slab}/(RA)\sub{bulk}$
\begin{align}
(RA)\sub{slab}/(RA)\sub{bulk}\approx \frac{1}{1+\alpha/t}
\label{eqn:RA_fit}
\end{align}
where $\alpha = (G\super{Nb}\sub{surf}+G\super{As}\sub{surf})/g\sub{bulk}$ (See Supplementary Note 3). For a material where the conduction is dominated by bulk states and the contribution of surface states is negligible, the value of $\alpha$ would be very small ($\alpha \rightarrow 0$). Conversely, materials like topological insulators, which exhibit zero bulk conductance and finite surface conductance would have $\alpha \rightarrow \infty$. Hence, the parameter $\alpha$ essentially quantifies where we are on the spectrum between a topological insulator and a conventional metal.

Equation~\ref{eqn:RA_fit} predicts that $(RA)\sub{slab} < (RA)\sub{bulk}$ in pristine slabs of any finite thickness $t$, as long as there is some surface contribution, $\alpha > 0$.
When surface conductance is negligible, $G\sub{surf}\rightarrow 0$ leading to $\alpha\to 0$, we find normalized $RA \to 1$ for all thicknesses, exactly as observed for conventional metals such as Cu. 
Agreement between the computed $(RA)\sub{slab}/(RA)\sub{bulk}$ and Equation~\ref{eqn:RA_fit} (Figure~\ref{fig:RA}(b)) establishes the validity of the simple model of additive surface and bulk conductance for Weyl semimetals.

We now investigate the effect of notches or surface line-defects on the ballistic conductance of NbAs films.
We study six different types of defects  as shown in Figure~\ref{fig:defects}(a-f).
The calculated transmission and the resultant normalized $RA$ product are shown in Figures~\ref{fig:defects}(g) and (h) respectively.
As expected, the transmission for pristine films increases linearly with thickness, which corresponds to the increasing number of bulk conducting channels/bands at the Fermi level.
We perform a linear fit ($y = mx+b$) for the thickness-dependent transmission data for all defect types.
(The parameters slope $m$, intercept $b$ and $R^2$ have been provided in Supplementary Table 1. We also use these parameters to compute the $RA$ product curves that would be expected if the transmission against thickness was a perfectly straight line as shown in Figure 5(h)).
Removing an As atom from the As-terminated (top) surface leads to a relatively small drop in the transmission for all the slabs, such that the intercept of the linear fit drops only slightly, $\Delta b \sim -0.6$.
However, removing a Nb atom from the Nb-terminated (bottom) surface causes an almost 4 times larger reduction ($\Delta b \sim -2.2$) in the transmission.
For the third case (Figure~\ref{fig:defects}(c)), where we remove an atom each from the top and bottom surface, the transmission reduces by $\sim 2.8$, which is equal to the sum of the above two reductions.
As we increase the depth of the `notch' on the surfaces (Figure~\ref{fig:defects}(d-f)), the net transmission continues to diminish.
We note that the total transmission extrapolated to zero thickness remains finite in films with single-atom, 2-atom, and 6-atom line-defects, indicative of the survival of surface-state conduction in films with sufficiently small disorder.
With the deep 12-atom and 20-atom defects, the transmission reduction levels off for the thinnest film and leads to the downturn in the $RA$ product with scaling as shown in Figure~\ref{fig:defects}(h).
This is similar to the resistivity scaling trend reported previously for CoSi films with high surface defect densities~\cite{lien2023unconventional}.

\begin{figure}[t!]
\includegraphics[width=\columnwidth]{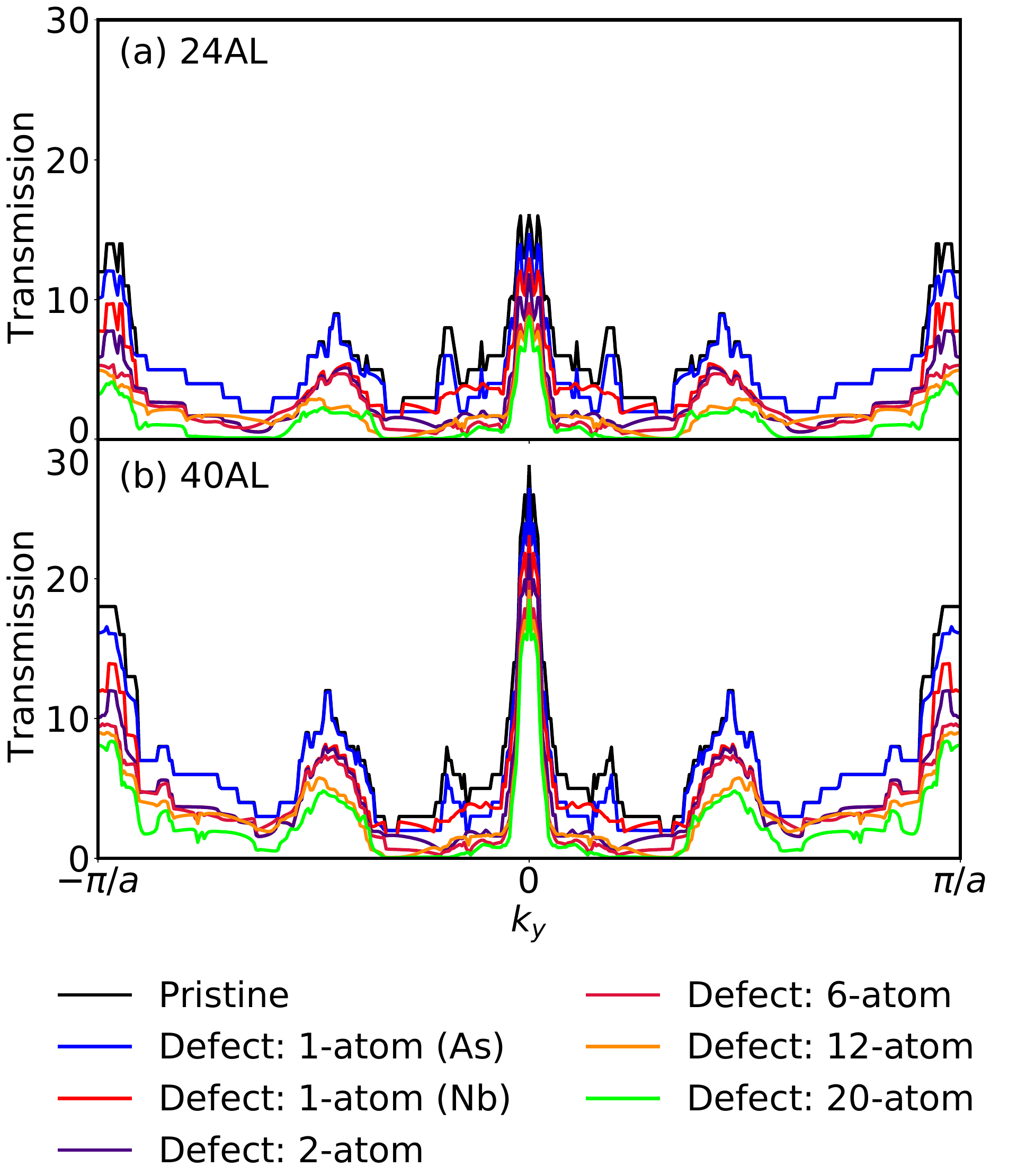}
\caption{\textbf{Momentum $\mathbf{k}$-resolved transmission of films of NbAs with surface line-defects} Increasing the thickness of the NbAs (001) slabs (results shown for 24AL and 40AL) increases the transmission (higher peaks) owing to the larger number of bulk states.
Except for the case of 1-atom (As) defect configuration, line defects reduce the transmission at every $\mathbf{k}$-point indicating the absence of protection.
For the 1-atom defect, where an atom is removed from the As-terminated surface, transmission is not affected in regions of the Brillouin zone which do not have the localized As-surface states.}
\label{fig:k_resolved} 
\end{figure}

To further understand the above observations, we analyze the $\mathbf{k}$-resolved transmission for two representative cases of 24AL and 40AL slabs. 
Figure~\ref{fig:k_resolved} shows the transmission plotted against direction $k_y$ which is the in-plane direction normal to the transport direction $k_x$.
Since the transmission for pristine films essentially represents the number of states at the Fermi level, the values are integers for any $k_y$.
As the thickness increases, we see an increase in the peak heights around $k_y\sim0$ , $k_y\sim \pm0.45\pi/a$, and $k_y\sim \pm\pi/a$ corresponding to the increasing number of bulk states around those points in the Brillouin Zone. (See Supplementary Figure 7).
In general, defects reduce the transmission, though by varying degree as noted in Figure~\ref{fig:defects}(g).
The 1-atom defect on the As-terminated surface negligibly changes the transmission for $k_y\sim[-0.9\pi/a,-0.47\pi/a]$ and $k_y\sim[0.47\pi/a,0.9\pi/a]$, because there are no surface states on the As-terminated surface in that region (Supplementary Figure 7).
Consequently, the localization of As-terminated surface states in the $\mathbf{k}$-space leads to the small overall change in transmission.
Most of the surface states at the Fermi level that contribute to conduction exist on the Nb-terminated surface, as shown previously in Figure~\ref{fig:conductance_contrib}.
Correspondingly, a defect on the Nb-terminated surface considerably reduces the transmission throughout $k_y$, since the Nb-terminated states extend throughout the projected 2D Brillouin Zone. 
The 2-atom defect, removing one Nb and one As atom on each surface, reduces the transmission by roughly the sum of the previous two cases.
We find the transmission reduces further for every point along $k_y$ for 6-atom, 12-atom and 20-atom defects.

Using the net transmission calculated above, we plot the normalized $RA$ product in Figure~\ref{fig:defects}(h) for the various defect configurations.
Since the transmission $T$ continues to exhibit a roughly linear dependence on thickness $t$ for the cases with defects, we could employ a model similar to Equation~\ref{eqn:RA_fit} to fit the computed data.
Specifically, we write $G = G_0T(\varepsilon_\mathrm{F}) = G_0 (mt+b)$.
Comparing to Equation~\ref{eqn:RA_fit}, we note that $\alpha = b/m$.
We find that the normalized $RA$ product for the first four defect types exhibit a trend similar to the pristine case, i.e. $(RA)\sub{slab}/(RA)\sub{bulk}$ decreases with decreasing thickness. Since the net transmission doesn't change significantly for the 1-atom defect on the As-terminated surface, its normalized $RA$ curve (blue) is very close to the case with no defects (black) in Figure~\ref{fig:defects}(h).
Increasing the size of the defects makes the $(RA)$ vs. $t$ curves flatter, as the normalized $RA$ product begins to become thickness independent and approaches 1 when the surface state conduction gradually diminishes, manifested in the intercept of transmission $b$ approaching zero. 
Thus, in the limit of sufficiently strong surface disorder, the Weyl semimetal behaves more like a conventional metal in this respect.
For example, the 12-atom and 20-atom defect configurations begin to kill the transmission of the bulk states besides significantly suppressing surface conduction, which makes the $RA$ product of the slab greater than that of the ideal bulk.

\begin{figure}[t!]
\includegraphics[width=0.9\columnwidth]{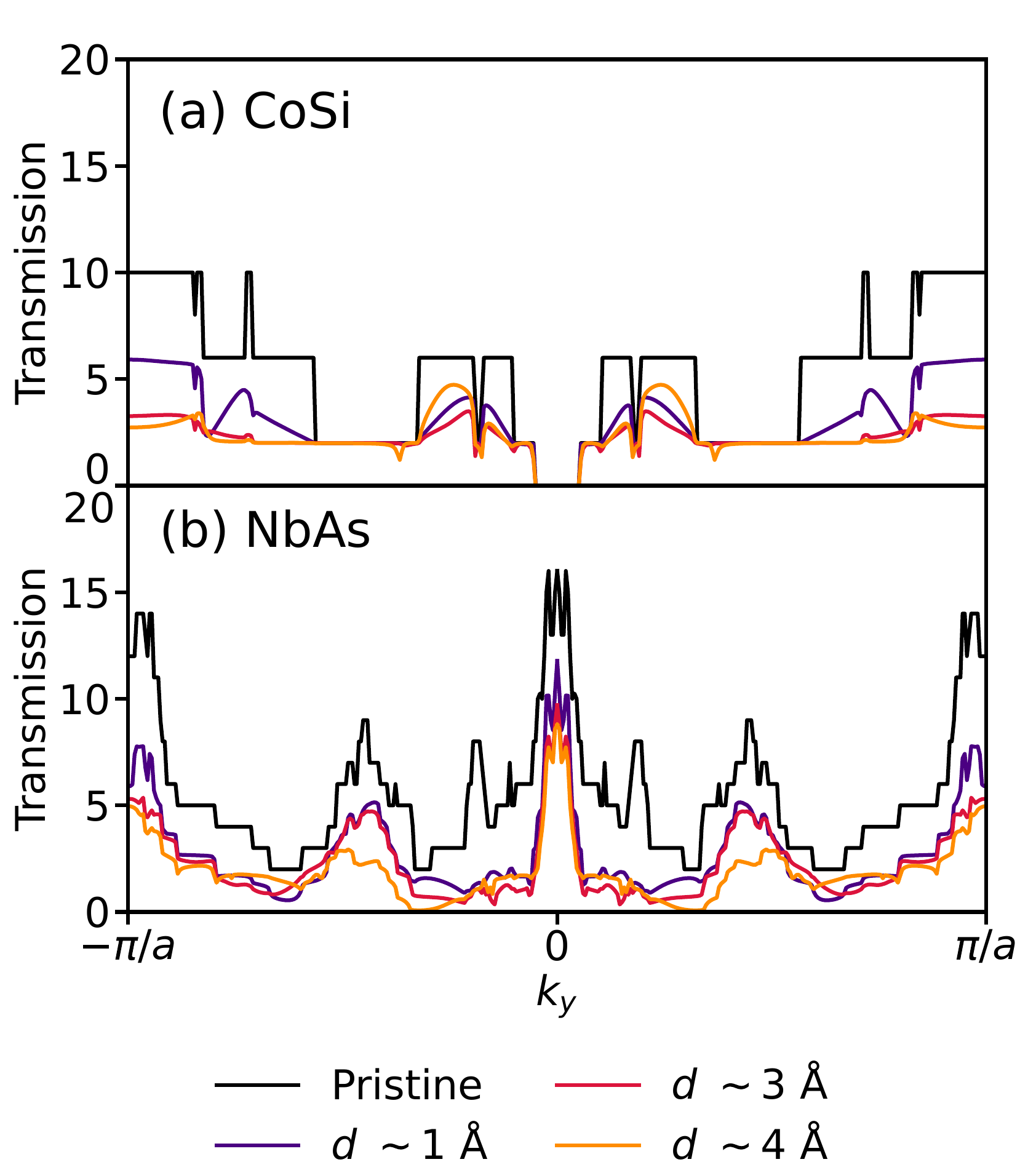}
\caption{\textbf{Comparison of the electron transport mechanism in CoSi and NbAs films} Momentum $\mathbf{k}$-resolved transmission for (001)-oriented slabs of (a) CoSi (3.3-nm thick)  and (b) NbAs (3.4-nm thick) in the presence of different surface line-defects. For the case of CoSi, there is one state per spin whose forward- and backward-moving surface states do not lie on the same surface due to chirality, and hence, their transmission never goes below 2 (accounting for spin degeneracy) as they cannot backscatter. However, the transmission is reduced throughout the Brillouin zone for NbAs since the forward- and backward-moving states lie on the same surface and so are not protected from backscattering.}
\label{fig:CoSi_NbAs_compare} 
\end{figure}

As a further proof that the transport in NbAs films is indeed dominated by surface states, we perform calculations for the cases where line-defects are introduced in the bulk (center) of the slabs (see Supplementary Figure 9(a-c)). Supplementary Figure 9(d) shows that the bulk line-defects do not have any significant impact on the normalized resistance-area product of the films. While a 2-atom surface defect reduces the transmission by around 44\% for a 4-nm slab (32AL), a 2-atom bulk defect decreases the corresponding transmission by only $\sim$ 5\%. This can be seen more clearly in the momentum-resolved transmission where the bulk line-defects have negligible effect on the transmission across the Brillouin zone (Supplementary Figure 10). These results reinforces our observation that the conduction is overwhelmingly dominated by surface states. Only surface defects significantly reduce the transmission and consequently, affect the resistance-area scaling.

It is important to note that for very thin films, the 12-atom and 20-atom defect configurations are large enough to significantly perturb the electronic states in the region near the defect.
In our current approach, however, the tight-binding models are based on the ground state of the pristine films, and the couplings linked to the removed atom are deleted to mimic the defect.
Therefore, the calculated transmissions at 16 AL and below for these two defect configurations are likely to be less reliable than the remaining cases.
Although more accurate results can be obtained using self-consistent DFT and NEGF in QuantumATK~\cite{smidstrup2020quantumatk}, it can be computationally expensive and potentially prohibitive for large-thickness structures with spin-orbit coupling, as studied here.
Nevertheless, the qualitative trend discussed above that the transmission levels off in the ultra-thin film limit is also demonstrated by the fully self-consistent DFT with NEGF calculations using QuantumATK~\cite{smidstrup2020quantumatk} (Supplementary Figure 1).

Though the transmission has been predicted for single line-defects in this work, the results can be extended to more realistic systems with multiple defects (The details of this approach can be found in the Supplementary Note 5). Supplementary Figure 2 shows that the resistivity of NbAs films in the diffusive limit decreases with thickness in the presence of weak surface disorder. This scaling trend of decreasing resistivity is similar to the resistance-area scaling observed for the films with single line-defects. However, as the surface transport is not well protected for NbAs, the resistivity first increases before decreasing as we increase the depth of the line-defects.

Finally, we compare the conductance scaling of NbAs films with that of CoSi.
Both materials show decreasing $RA$ with reduced thickness in pristine films, owing to the dominance of surface conduction over bulk conduction at the nanometer scale.
However, CoSi is a chiral semimetal with forward- and backward-moving surface carriers from the Fermi-arc states of the same transverse momentum spatially separated on opposite surfaces of a CoSi thin film~\cite{chen2020topological}.
Consequently, line defects which preserve the transverse momentum cannot backscatter these states into each other and the transmission of the CoSi Fermi arc states is robust against such defects~\cite{chen2020topological}.
In contrast, the forward- and backward-moving surface carriers with the same transverse momentum coexist on both surfaces of NbAs, and thus can intermix (see Supplementary Figure 11).
Therefore, transmission of the NbAs Fermi-arc states is much more susceptible to defect scattering.
As shown in Figure~\ref{fig:CoSi_NbAs_compare}, a line-defect can reduce the transmission for all $\mathbf{k}$-points in the Brillouin zone, contrary to that in the CoSi films.
This explains the substantial reduction of total transmission in NbAs films with single-atom defects (see Figure~\ref{fig:defects}). Supplementary Figure 14 shows the NEGF-predicted transmissions and resultant $RA$ products for the two materials. To make a fair comparison, we introduce surface line-defects of comparable depths. Unlike in the case of CoSi where a decreasing $RA$ trend persists for deeper defects, in NbAs, we begin to 
observe a near-reversal (i.e., $RA$ product starts to increase with decreasing thickness) for a 4-$\textup{\AA}$-deep notch. This is consistent with our observation above that the transport is more robust to surface defects in CoSi (a chiral topological semimetal) than in NbAs. We also note that though the $RA$ product begins to flatten out earlier for NbAs, the absolute value of the $RA$ product is still lower (i.e., higher conductance) than that of CoSi, because NbAs has a higher number of conducting surface states at the Fermi level than CoSi.

Experimentally, the 2-3 orders of magnitude increase in resistivity observed in $\sim$600 nm diameter NbAs nanowires~\cite{xu2015discoveryNbAs} when compared to the micron-size wide nanobelts demonstrates the sensitivity of the NbAs surface states to defect scattering at the boundaries.
Therefore, for materials with optimal disorder-tolerant surface-state conductivity, future work should explore \emph{chiral} topological semimetals. In addition, while both NbAs and CoSi beat Cu in terms of the normalized-$RA$ product (i.e., $RA_{\mathrm{Slab}}/RA_{\mathrm{Bulk}}$) for thinner films, in absolute terms the $RA$ product of Cu is significantly lower than that of either of the two topological semimetals in pristine films because Cu has many more conducting states at the Fermi level to start with. Hence, even with the loss of transmission in the presence of defects, Cu may still outperform CoSi and NbAs. Therefore, it’s imperative to search for topological semimetals that are both chiral in their structure and also have large numbers of Fermi arcs to maximize the conducting chiral surface states. 

\section*{Discussion}

In summary, we performed first-principles NEGF calculations to understand the mechanism of electron transport in thin films of a representative Weyl semimetal, NbAs.
The resistance-area $RA$ product in pristine NbAs films decreases with thickness at the nanometer scale, in contrast to a nearly constant $RA$ product in ideal Cu films.
This anomalous scaling is the manifestation of the numerous surface states in the bandstructure of NbAs. The surface states account for over 70$\%$ of the conductance for 2.1-nm-thick (relaxed 16 AL) films and $\sim$ 50$\%$ for 6.8-nm-thick (relaxed 48 AL) films; furthermore, contribution from the Nb-terminated surface states is almost 3 times that of the As-terminated-surface states.
The decreasing $RA$ with reducing dimensions persists even with surface defects, as long as the degree of disorder is moderate.
This contrasts the ever increasing $RA$ with reducing dimensions in conventional metals like Cu when disorder is present, and highlights the promise of Weyl semimetals, and topological semimetals in general, for integrated circuits. Finally, analyses of electron transmission in $\mathbf{k}$-space show that electron transport in NbAs is not immune to defect scattering because forward- and backward-moving states coexist on the same surface, in contrast to the protected chiral surface transport in CoSi thin films.
The comparison between the two material systems calls for the search for \emph{chiral} topological semimetals with large numbers of Fermi arcs for low-resistance nanoscale interconnects. 

\section*{Methods}

\subsection*{First-principles calculations}
We use open-source plane-wave DFT software JDFTx~\cite{JDFTx} for the generation of self-consistent relaxed crystal structures, electron bandstructures and Wannier tight-binding models.
We use the fully-relativistic optimized norm-conserving Vanderbilt pseudopotentials (ONCVPSP)~\cite{hamann2013optimized} as distributed by the open-source PSEUDODOJO library~\cite{van2018pseudodojo} to include spin-orbit coupling self-consistently.
These DFT calculations are performed using the Perdew-Burke-Ernzerhof (PBE) generalized gradient approximation (GGA) to the exchange-correlation functional~\cite{Perdew} at a plane-wave cutoff of 40 Hartrees and a charge density cutoff of 200 Hartrees.

For the first-principles study of NbAs slabs, we construct films of (001) orientation to allow direct comparison of our computed Fermi surfaces with the available ARPES results, which have been experimentally measured for the cleaved (001) surfaces~\cite{xu2015discovery,sun2015topological}.
These slabs have tetragonal unit cells, and are constructed with a vacuum spacing of $12\textup{~\AA}$ thickness along the $c-$direction, employing Coulomb truncation to eliminate long-range interactions between periodic images along this direction~\cite{sundararaman2013regularization, ismail2006truncation, rozzi2006exact}.
Cleaving the surface along (001) direction leads to two asymmetric surfaces with Nb and As terminations respectively (Figure~\ref{fig:intro}(a)), which produces an overall dipole moment in the unit cell.
Supplementary Figure 3 shows that the Coulomb truncation scheme accounts for this dipole correctly and produces zero electric field in the vacuum region away from both surfaces.

With the computational setup described above, we first perform an optimization of the ionic positions and lattice parameters of the body-centered tetragonal unit cell of bulk NbAs (space group $I4_1md$).
The initial crystal structure was obtained from the Materials Project database~\cite{Jain2013}.
The relaxation yields lattice constants of $a = b = 3.46 \textup{~\AA}$ and $c = 11.80 \textup{~\AA}$ which are within $\sim$ 1\% of the XRD measured values of  $a = 3.45 \textup{~\AA}$ and $c = 11.68 \textup{~\AA}$ (Figure~\ref{fig:intro}(a))~\cite{xu2015discovery,boller1963transposition}.
Starting from a single-unit-cell thick slab, we then construct films with seven different thicknesses in steps of 1 unit cell.
Hence, the thickness of our films vary from 1 unit cell or 8 atomic layers (AL) to 7 unit cells or 56 atomic layers (AL).
Previous first-principles calculations for NbAs have found no noticeable change in the band structure and Fermi surfaces for slabs larger than 7 unit cells in thickness~\cite{sun2015topological}. The DFT calculations for the bulk and slabs are performed using \textbf{k}-point meshes of $8\times8\times2$ and $8\times8\times1$ respectively, and Fermi smearing with width 0.01~Hartrees.
Keeping the in-plane lattice constants of the slabs fixed ($a=b$), we optimize the ionic positions using self-consistent DFT for subsequent calculations of electronic bands, Fermi surface and electron transport properties.

\subsection*{Creating tight-binding models}

We then construct a tight-binding model using a maximally-localized Wannier function basis set~\cite{WannMarzari} in JDFTx.
Supplementary Figure 3 shows the contribution of $s-$, $p-$ and $d-$orbitals of Nb and As atoms to each band for a 16AL slab.
The electron bands in the energy range $\pm 6.5$ eV around the Fermi level are mostly composed of the $d-$ and $p-$orbitals of Nb and As atoms respectively.
Hence, we choose a basis set of 10 $d-$orbitals per Nb atom and 6 $p-$orbitals per As atom in the unit cell as the initial guesses.
We construct maximally-localized Wannier functions for our \emph{ab initio} tight-binding model that reproduces the DFT bands in the energy window of $\varepsilon_\mathrm{F} - \sim 7.3$ eV to $\varepsilon_\mathrm{F} + \sim2.9$ eV above $\varepsilon_\mathrm{F}$, as shown in Supplementary Figure 5.

\subsection*{Calculating surface and bulk contributions}

To pinpoint the contributions of surface and bulk contributions to the band structure, Fermi surfaces and conductance in the Wannier basis, we compute weights of each Wannier-interpolated electronic state from the surface regions.
Specifically, we define functions $w^X(z)$ for $X$ = Nb and As, which are 1 within the dashed rectangles shown in the bottom panel of Fig.~\ref{fig:conductance_contrib}, and 0 outside it.
We then compute the matrix elements $w^X_{\vec{k}ab} \equiv \int_\Omega \mathrm{d}\vec{r} \psi_{\vec{k}a}^\ast(\vec{r}) \bar{w}^X(z) \psi_{\vec{k}b}(\vec{r})$,
where $\bar{w}^X(z)$ is $w^X(z)$ smoothed by convolution with a Gaussian of width 1 bohr.
Finally, we interpolate $w^X_{\vec{k}ab}$ using the Wannier representation in exactly the same way as the Hamiltonian and momentum matrix elements described in detail elsewhere~\cite{habib2018hot, kumar2022fermi}.

\subsection*{Non-equilibrium Green’s Function calculations}

Using the tight-binding models created above, we employ  Non-equilibrium Green’s Function (NEGF) method to compute the electron transport properties of the films~\cite{datta2005quantum}. For the slab of NbAs, we consider transport along the [100] direction and calculate the total transmission as 
\begin{align}
    T(E) = \int d k_y T(k_y,E)
    \label{eqn:T_slab}
\end{align}
where $T(k_y,E)=Tr(\Gamma_L G^R \Gamma_R G^A)$ is the $k_y$-resolved transmission and $k_y$ is the in-plane direction.
Here, $G^R(k_y,E)=[E+i\eta-H_{C,k_y}-\Sigma(k_y,E)]$ is the retarded Green’s function, $H_{C,k_y}$ is the tight-binding Hamiltonian of channel, and $\Sigma(k_y,E)=\Sigma_L(k_y,E)+\Sigma_R(k_y,E)$ is the contact self-energy of the left $(L)$ and right $(R)$ contact. $G^A(k_y,E)$ is the advanced Green’s function, and $\Gamma_\alpha=i(\Sigma_\alpha-\Sigma_\alpha^\dagger)$ is the broadening of the contact-$\alpha$ $(\alpha=L,R)$. The contact self-energies are numerically solved using the Sancho-Rubio’s method~\cite{sancho1984quick}. For the bulk of NbAs, similarly, the transmission can be written as
\begin{align}
    T(E) = \int d k_y d k_z T(k_y,k_z,E)
    \label{eqn:T_bulk}
\end{align}
where the $k_z$ is the out-of-plane direction for the bulk. We use $k$-point sampling of $400$ $k_y$ and $800$ $k_y\times 800$ $k_z$ for slab and bulk transport calculation, respectively.
The Hamiltonian of the channel $H_C$ is constructed from the slab tight-binding model.
In order to consider surface defect configurations in the channel, we remove the orbitals of atoms entirely from the Hamiltonian of channel $H_C$.
Supplementary Figure 6 shows the schematic view of structure for NEGF calculation for 24AL slab of NbAs with 12-atom defect configuration. 

We, however, note that the method described here would be too expensive for predicting how impurity scattering in the form of point defects could affect the scaling trends. One could use the method recently proposed by Lien et al.~\cite{lien2023unconventional} for this purpose.
Additionally, our study does not include the effect of electron-phonon scattering which is beyond the scope of this work and is an important direction for future studies.

\section*{Data availability}

All relevant data are available from the authors upon request.

\section*{Code availability}

First-principles methodologies available through open-source software, JDFTx, and post-processing scripts available from authors upon request.

\section*{Acknowledgements}

S.K and R.S. acknowledge funding from Semiconductor Research Corporation under Task No. 2966.002.
Calculations were carried out at the Center for Computational Innovations at Rensselaer Polytechnic Institute. The work at the National University of Singapore was supported by MOE-2017-T2-2-114, MOE-2019-T2-2-215, and FRC-A-8000194-01-00. T.-R.C. was supported by 2030 Cross-Generation Young Scholars Program from the Science and Technology Council (MOST111-2628-M-006-003-MY3), Cheng Kung University, and the Center for Theoretical Sciences. We gratefully acknowledge the helpful discussions with Daniel Gall (RPI), Utkarsh Bajpai (IBM) and Vijay Narayanan (IBM). 


\section*{Author contributions}

S.K. and Y.-H.T. performed the first-principles calculations and analyzed the data. S.K. and C.-T.C. wrote the manuscript with input from all the authors. C.-T.C. conceived and supervised the project.

\section*{Competing Interests}

The authors declare no competing interests.

\bibliography{references}

\end{document}


\singlespacing
\maketitle

\setcounter{equation}{0}
\setcounter{figure}{0}
\setcounter{table}{0}
\setcounter{page}{1}
\makeatletter

\section{Supplementary Note 1: Ballistic electron transport in a 1D conductor with one band}

Consider a 1D conductor that is contacted at both ends by an external circuit. At the left contact, the current due to electrons moving in the right direction is~\cite{Rana}:
\begin{align}
    I_{L\longrightarrow R} = (\mathrm{-e})2 \times \int_0 ^\infty \frac{dk_x}{2\pi}v(k_x)f(E(k_x)-E_{fL}) \label{sum}
\end{align}

Similarly, at the right contact, the current due to electrons moving in the left direction is:
\begin{align}
    I_{R\longrightarrow L} = (\mathrm{-e})2 \times \int_{-\infty}^0 \frac{dk_x}{2\pi}v(k_x)f(E(k_x)-E_{fR})
\end{align}

The net current is the sum of the currents due to the right-moving and left-moving electrons:
\begin{align}
    I &= I_{L\longrightarrow R}+ I_{R\longrightarrow L}\\
    &= (\mathrm{-e})2 \times \int_0 ^\infty \frac{dk_x}{2\pi}v(k_x)f(E(k_x)-E_{fL}) + (\mathrm{-e})2 \times \int_{-\infty}^0 \frac{dk_x}{2\pi}v(k_x)f(E(k_x)-E_{fR}) \\
    & = 2e \times \int_0 ^\infty \frac{dk_x}{2\pi}v(k_x)[f(E(k_x)-E_{fR})- f(E(k_x)-E_{fL})] \\
    &=  2e \times \int_0 ^\infty \frac{dk_x}{2\pi}v(k_x)\left[ f(E(k_x))-E_{fR} \times \frac{\partial f(E(k_x))}{\partial E(k_x)}- f(E(k_x))+E_{fL} \times \frac{\partial f(E(k_x))}{\partial E(k_x)}\right] \\
    &= 2e \times \int_0 ^\infty \frac{dk_x}{2\pi}v(k_x) (E_{fL}-E_{fR}) \times \frac{\partial f(E(k_x))}{\partial E(k_x)}
\end{align}
For electrons, $E_{fL}-E_{fR}$ is the potential difference across the conductor which could be written as $eV$. Since $G = \partial I / \partial V$, the conductance $G$ can be written as:
\begin{align}
    G =2e^2 \times \int_0 ^\infty \frac{dk_x}{2\pi}v(k_x)  \frac{\partial f(E(k_x))}{\partial E(k_x)}
\end{align}

\section{Supplementary Note 2: Conductance for a general 3D conductor}

For simplicity, let's assume that the current is flowing in the $x-$direction. For a more general case where an electronic state can have both positive and negative velocities irrespective of the sign of the $\vec{k}-$point, the summation in equation \ref{sum} is performed over whether the electrons are moving in $+x$ or $-x$ direction. Hence, $I$ becomes:
\begin{align}
    I &= I_{L\longrightarrow R}+ I_{R\longrightarrow L}\\
    &= (\mathrm{-e})\int_{BZ} g_s\frac{d\vec{k}}{(2\pi)^3} \left[
    v^x_{\vec{k}n}  \Theta (v^x_{\vec{k}n}) f(\varepsilon_{\vec{k}n}-\varepsilon_{fL})
    + v^x_{\vec{k}n}  \Theta (-v^x_{\vec{k}n}) f(\varepsilon_{\vec{k}n}-\varepsilon_{fR})\right] \\
    &= (\mathrm{-e})\int_{BZ} g_s\frac{d\vec{k}}{(2\pi)^3} \left[
    v^x_{\vec{k}n}  \Theta (v^x_{\vec{k}n}) f(\varepsilon_{\vec{k}n}-\varepsilon_{f}+(\mathrm{-e})V/2)
    + v^x_{\vec{k}n}  \Theta (-v^x_{\vec{k}n}) f(\varepsilon_{\vec{k}n}-\varepsilon_{f}-(\mathrm{-e})V/2)\right] \\
\end{align}
expressing the reservoir potentials in terms of their mean $\varepsilon_f$ and a potential difference $V$. Here, $\Theta(v^x_{\vec{k}n})$ is the heavy side step function which is 1 when $v^x_{\vec{k}n} >0$ and zero otherwise. Note, $g_s$ is the spin-degeneracy factor, which would be 1 for DFT calculations with spin orbit coupling. Taylor expanding about the Fermi distributions at the mean $\varepsilon_f$ (denoted $f_{\vec{k}n}$):
\begin{align}
I &= (\mathrm{-e})\int_{BZ} g_s\frac{d\vec{k}}{(2\pi)^3} \left[
    v^x_{\vec{k}n}  \Theta (v^x_{\vec{k}n}) (f_{\vec{k}n} - (eV/2) f'_{\vec{k}n})
    + v^x_{\vec{k}n}  \Theta (-v^x_{\vec{k}n}) (f_{\vec{k}n} + (eV/2) f'_{\vec{k}n})\right] \\
&= (\mathrm{-e})\int_{BZ} g_s\frac{d\vec{k}}{(2\pi)^3} \left[
\begin{array}{c}
f_{\vec{k}n} (v^x_{\vec{k}n} \Theta (v^x_{\vec{k}n}) + v^x_{\vec{k}n} \Theta (-v^x_{\vec{k}n}))\\
+ (-eV/2) f'_{\vec{k}n} (v^x_{\vec{k}n} \Theta (v^x_{\vec{k}n}) - v^x_{\vec{k}n} \Theta (-v^x_{\vec{k}n}))
\end{array}\right] \\
&= (\mathrm{-e})\int_{BZ} g_s\frac{d\vec{k}}{(2\pi)^3} \left[
    f_{\vec{k}n} v^x_{\vec{k}n} + (-eV/2)f'_{\vec{k}n} |v^x_{\vec{k}n}|) \right] + O(V^2)
\end{align}

Note, 
\begin{align}
    \Theta (v^x_{\vec{k}n}) = \frac{v^x_{\vec{k}n}+|v^x_{\vec{k}n}|}{2v^x_{\vec{k}n}}
\end{align}
First term is current at equilibrium, which should be zero.
Conductance is therefore:
\begin{equation}
G = \frac{\partial I}{\partial V}
= \frac{e^2}{2}\int_{BZ} g_s\frac{d\vec{k}}{(2\pi)^3} f'_{\vec{k}n}|v^x_{\vec{k}n}|
\end{equation}

The derivative of Fermi function $f'_{\vec{k}n}$ is implemented in the code as: 
\begin{equation}
f'_{\vec{k}n} = \frac{\beta}{ 4 \cosh^2 \left[ \beta/2\left( \varepsilon_{\vec{k}n}-\mu\right) \right] }\notag
\end{equation}

where $\beta = 1/kT$ and is essentially the smearing width in energy units. 

\section{Supplementary Note 3: Analytical expression for ballistic conductance for pristine NbAs slabs}

We can write the total conductance of a NbAs slab $G$ as 

\begin{align}
    G &= g_{\mathrm{Bulk}} t + G^{\mathrm{Nb}}_{\mathrm{Surf}}+G^{\mathrm{As}}_{\mathrm{Surf}}\\
    &=g_{\mathrm{Bulk}} t + G_0
\end{align}
where $t$ is the thickness of the slab, $g_{\mathrm{Bulk}}$ is the slope of the linear relationship between $G$ and $t$, and  $G^{\mathrm{Nb}}_{\mathrm{Surf}}$ and $G^{\mathrm{As}}_{\mathrm{Surf}}$ are the thickness-independent contributions of the Nb-terminated- and As-terminated-surface states respectively. We can club together the surface conductance contributions into one term $G_0$, such that $G_0 = G^{\mathrm{Nb}}_{\mathrm{Surf}}+G^{\mathrm{As}}_{\mathrm{Surf}}$. The $RA$ product for a slab, thus, becomes

\begin{align}
    (RA)_{\mathrm{Slab}} = \frac{A}{G}= \frac{A}{g_{\mathrm{Bulk}} t + G_0}
\end{align}
where the cross-sectional area $A$ of a film of thickness $t$ and in-plance lattice constant $a$ would be $ta$. Hence,

\begin{align}
    (RA)_{\mathrm{Slab}}&= \frac{at}{g_{\mathrm{Bulk}} t + G_0}\\
     &= \frac{a}{g_{\mathrm{Bulk}} + G_0/t}
\end{align}

For bulk, $t \rightarrow \infty$,
\begin{align}
    (RA)_{\mathrm{Bulk}}= \frac{a}{g_{\mathrm{Bulk}}}
\end{align}

Combining the above two equations,
\begin{align}
    (RA)_{\mathrm{Slab}}/(RA)_{\mathrm{Bulk}}&= \frac{g_{\mathrm{Bulk}}}{a} \cdot \frac{a}{g_{\mathrm{Bulk}} + G_0/t} \\
    &= \frac{1}{1+G_0/g_{\mathrm{Bulk}}\cdot 1/t}\\
    &= \frac{1}{1+\alpha/t}
\end{align}

where $\alpha =G_0/g_{\mathrm{Bulk}}$

\section{Supplementary Note 4: Validation using fully self-consistent density-functional Non-Equilibrium Green's Function (NEGF) method}

In our work, we use the Wannier tight-binding models generated using density-functional theory (DFT) to predict the transmission of the slabs of varying thickness. In order to mimic the line defects, we delete the orbitals linked with the atoms we intend to remove. To validate this approach, we use fully self-consistent DFT framework as implemented in QuantumATK~\cite{smidstrup2020quantumatk,brandbyge2002density}. We perform the simulations for 8AL, 16AL, 24AL and 32AL  slabs of (001) orientation of NbAs.  We use double-zeta-polarized basis set with a cut off of 45 Hartrees and a Fermi smearing of 0.01 Hartrees. The NEGF device consists of left and right terminals and a central scattering region such that the direction of transport is along the $z$-direction with $y$-direction being the in-plane periodic direction. A vacuum of around $\sim17 \textup{~\AA}$ along the $x$-direction i.e. perpendicular to the (001) surface is set. For our calculations, we employ a $\mathbf{k}$-point sampling of $8\times8\times271$. The channel/scattering region has a length of around $60\textup{~\AA}$ ($\sim80\textup{~\AA}$ for 12-atom and 20-atom defect configurations) and the left and right electrodes are each $13.94 \textup{~\AA}$ long. 

For the purpose of validation, we are only looking for agreement in the qualitative trends between the two methods. This is due to two reasons: (1) The approach of deleting the Hamiltonian matrix elements associated the deleted atoms might not accurately represent the electronic state near the defects and (2) since QuantumATK performs electronic structure calculations for a large system of atoms, calculations are generally performed with the linear combination of atomic orbits (LCAO) method using a basis set, which is faster but less accurate than the planewave DFT approach used for generating Wannier tight-binding models in JDFTx. Hence, we expect some discrepancies between the bandstructure and Fermi surfaces derived from the two methods, which would quantitatively affect the transmission results. However, the qualitative trends should be consistent.   

\begin{figure}[htp!]
\centering
\includegraphics[width=\textwidth]{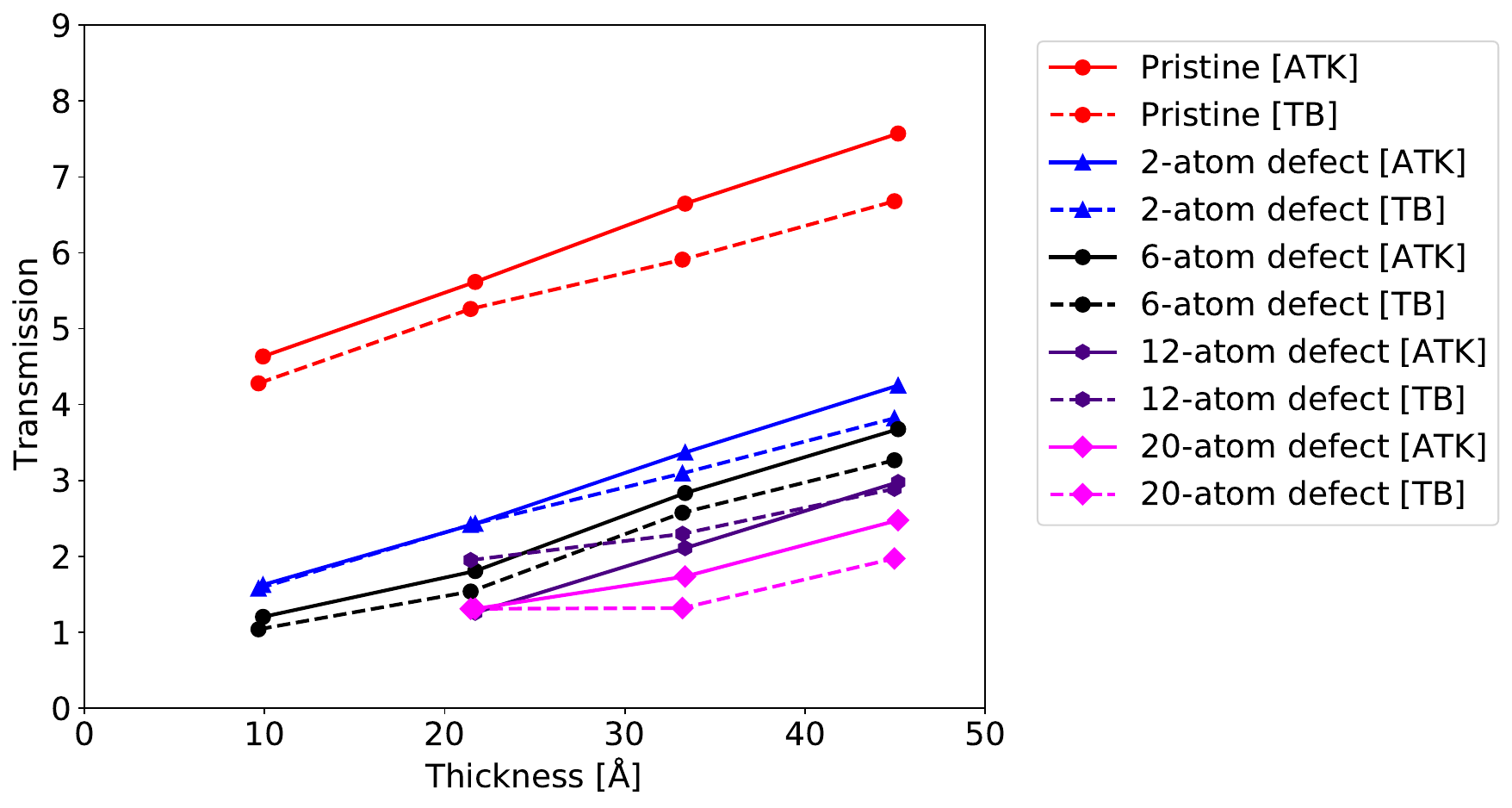}

\caption{Comparison of the transmissions computed using the in-house NEGF code with Wannier tight-binding (TB) models (dashed lines) and QuantumATK (solid lines) for 8AL, 16AL, 24AL and 32AL films with and without defects.}
\label{fig:ATK} 
\end{figure}

We perform the calculations in QuantumATK without spin-orbit coupling (SOC) since calculations with SOC would be computationally very expensive for large slab thickness. For comparison with the method used in this paper, we regenerate the tight-binding models using non-relativistic pseudopotentials in JDFTx and calculate the transmission using the NEGF method. Supplementary Figure~\ref{fig:ATK} shows the comparison of the transmissions calculated using ATK and our method for different defect configurations. We see very good agreement between the two with the transmission values lying within $12\%$ of each other for most cases. Significant deviations are observed for line defects of larger depth (e.g. 12-atom defect).

\section{Supplementary Note 5: Resistivity scaling for films with multiple line-defects}

In our work, we compute the transmission and resultant conductance (or resistance) of NbAs films with single line-defects of varying sizes and types (as shown in Figure 5). While these results are valid for a single line-defect (i.e., a single scatterer), we can easily extend these results to predict the resistance of films with a given linear defect density $\nu$ of a particular defect configuration. Specifically, the transmission probability $\tau(L)$ of a film of length $L$ and thickness $t$ with $N$ line-defects in series (where, $N = \nu L$) can be written as \cite{datta1997electronic}: 
\begin{equation}
    \tau(L,t)=\frac{L_0}{L+L_0}
\end{equation}
where
\begin{align}
    L_0(\nu,t) = \frac{\tau(t)}{\nu(1-\tau(t))}
\end{align}
Here, $\tau(t)$ is the transmission probability of a single line-defect (i.e., single scatterer) of a film of thickness $t$ computed using our first-principles framework. The transmission $T$ would therefore be $\tau M$, where $M$ is the number modes in the system. The value of $M$ is essentially the transmission of the pristine films. Following the above equations, the conductance as a function of length $L$ and thickness $t$ is given as:
\begin{align}
    G(L,t) = G_0M(t) \tau(L,t)=\sigma(L,t)\frac{A}{L}
\end{align}
where $G_0$ is the conductance quantum ($G_0 = 2e^2/h$). The conductivity $\sigma(L,t)$ hence becomes:
\begin{align}
    \sigma(L,t)=G_0M(t)\tau(L,t)\frac{L}{A}=\frac{G_0M(t)}{A}\frac{LL_0(\nu, t)}{L + L_0(\nu,t)}
\end{align}
The conductivity in the diffusive limit ($L>>L_0$) would be
\begin{align}
    \sigma(L>>L_0,t) = \frac{G_0M(t)}{A}L_0(\nu,t)
    \label{diffusive_sigma}
\end{align}
We can use the above formalism to calculate the resistivity of films due to defect scattering Note that this formalism is only valid for uncorrelated defects, so strictly speaking, the calculations are reliable only for samples with sufficiently low defect densities. We therefore limit our studies to films with $\nu \le $ 0.05 (i.e., 5 line-defects per 100 unit cells).

\begin{figure}[htp!]
\includegraphics[width=\textwidth]{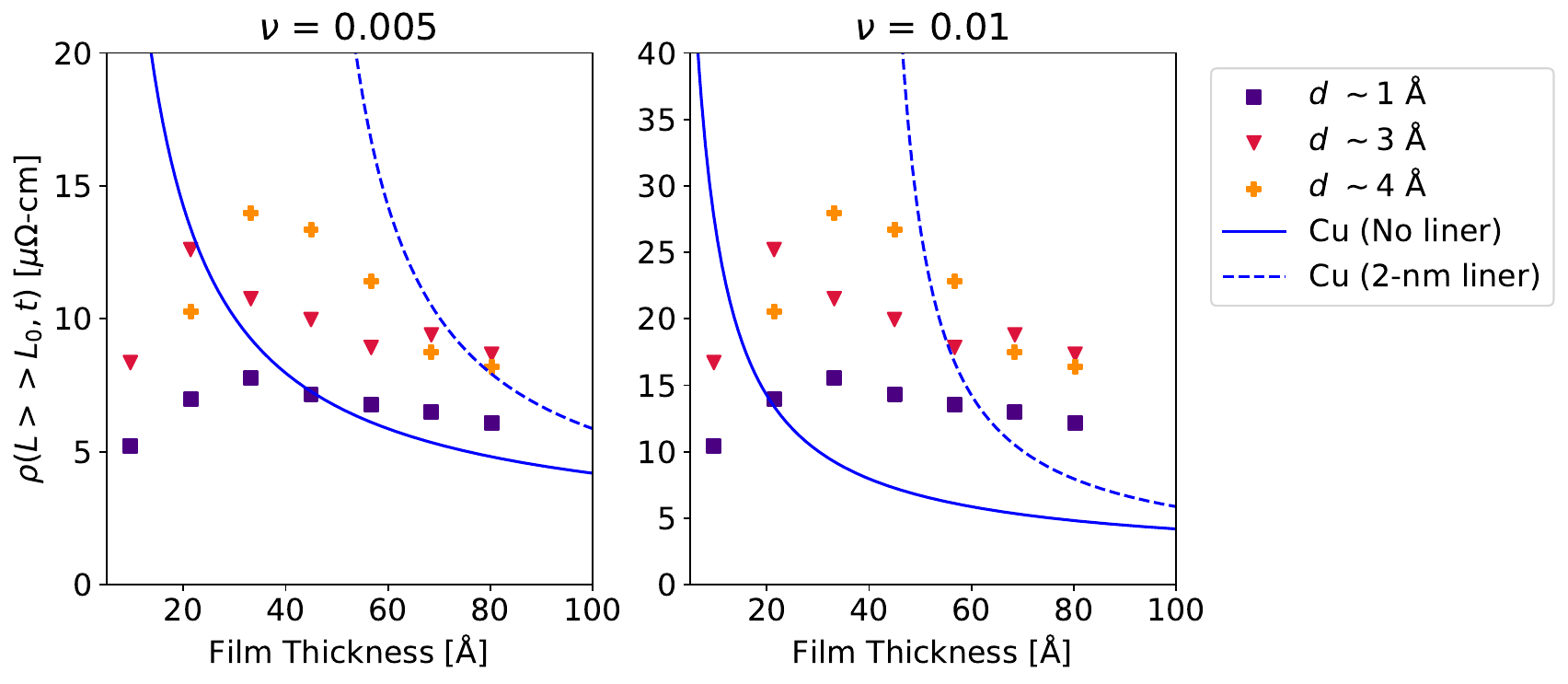}
\caption{Resistivity scaling for (001) NbAs films with defect densities of (a) 0.005 and (b) 0.01 in the diffusive limit calculated using the transmission probability of a single line-defect (or scatterer). The blue lines are the resistivities calculated for Cu films (with and without liners) using the Fuchs-Sondheimer model. The calculations assume completely diffuse scattering at the film surfaces and models the electron transport in films without bulk defect or grain boundary scattering.}
\label{fig:line_density} 
\end{figure}

Supplementary Figure~\ref{fig:line_density} shows the impact of a linear density of 0.005 (i.e., 5 line-defects per 1000 unit cells) and 0.01 (i.e., 10 line-defects per 1000 unit cells) on the resistivity of NbAs films as a function of thickness for different defect configurations. We see that the resistivity of NbAs films in the diffusive limit decreases with thickness in the presence of weak surface disorder. As the surface transport is not well protected for NbAs, the resistivity first increases before decreasing as we increase the depth of the line-defects. 

In a defect-dominated transport regime, the scaling trend for resistivity in the diffusive limit is similar to the resistance-area scaling observed for the single line-defects earlier. Specifically, if $a$ is the lattice constant, $t$ is the thickness of the slab, then $RA$ product would be:
\begin{align}
    RA = \frac{A}{G} = \frac{a t}{G_0 T(t)}=\frac{a t}{G_0M(t)\tau(t)}=\frac{a  t}{G_0(c_1t+c_2)\tau(t)}
    \label{RA}
\end{align}
In the above equation, $M(t)$ has been expressed as a linear function of thickness $t$ with constants $c_1$ and $c_2$, which denote the slope and intercept of the line respectively. This linear relationship between $M(t)$, which is the transmission of the slab under pristine conditions and thickness $t$ has been demonstrated by our first-principles calculations (Figure 4). Physically, it implies that the number of modes is directly proportional to the thickness of the slabs and has a constant, non-zero surface contribution even as $t\rightarrow 0$. For comparison, the resistivity as a function of thickness in the diffusive limit (\ref{diffusive_sigma}) can be simplified to:
\begin{align}
    \rho(L>>L_0,t)=\frac{A}{G_0M(t)L_0}=\frac{at}{G_0(c_1t+c_2)}\frac{\nu(1-\tau(t))}{\tau(t)}
    \label{diffusive_rho}
\end{align}
Equations \ref{RA} and \ref{diffusive_rho} show the difference in the scaling trend between the $RA$ product and resistivity $\rho(L>>L_0)$ in the diffusive limit.

To compare this to Cu, which is the most widely used interconnect metal, we plot the resistivity of thin films of Cu as a function of film thickness. We use the Fuchs-Sondheimer model—a semi-classical model used to predict the increase in resistivity due to size effects for conventional metallic films~\cite{fuchs1938math,sondheimer1952adv}. The model assumes that Cu films have no bulk defects or grain boundaries and hence, represent the best-case scenario for Cu. The results show that below 4 nm, the resistivity of NbAs is considerably lower than that of Cu (up to an order of magnitude) which highlights the potential of NbAs as an interconnect metal.  Furthermore, due to its low resistance to electromigration effect, Cu requires a thin liner (up to 2-nm thick)~\cite{gall2020search} which greatly exacerbates its performance as interconnects. Our first-principles calculations predict that the cohesive energy per atom of NbAs is 6.2 eV/atom, which is around 1.8 times that of Cu (= 3.4 eV/atom). This suggests that NbAs potentially doesn’t need a liner due to its relative stability, which gives it a significant advantage over Cu. Assuming that NbAs interconnects would be liner-free, NbAs beats Cu below 8 nm for 0.5\% defect density for all defect configurations in the limit of disorder-dominated transport. For a 1\% defect density, NbAs outperforms Cu below 6 nm for the different defects studied here.

Our calculations presented here also demonstrate that this approach can be extended to more realistic systems with larger defect densities.

\newpage
\begin{figure}[htp!]
\includegraphics[width=\textwidth]{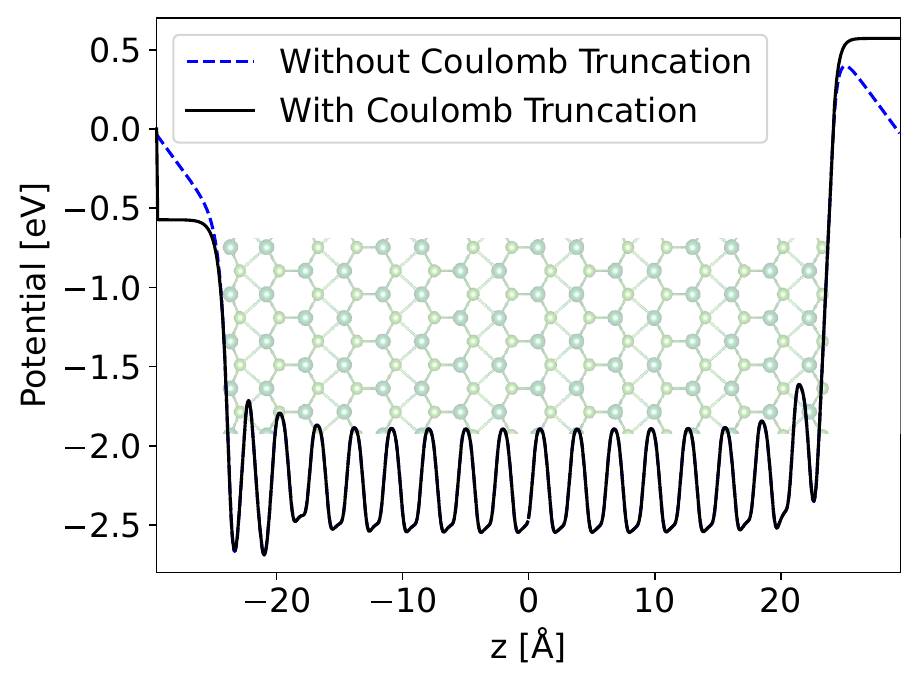}
\caption{Electrostatic potential in the film and vacuum region of a 32AL (001)-terminated NbAs slab ($\sim 44.96 \textup{~\AA}$) for the cases of (i) with and (ii) without Coulomb truncation.}
\label{fig:potential} 
\end{figure}

\newpage
\begin{figure}[htp!]
\includegraphics[width=0.8\textwidth]{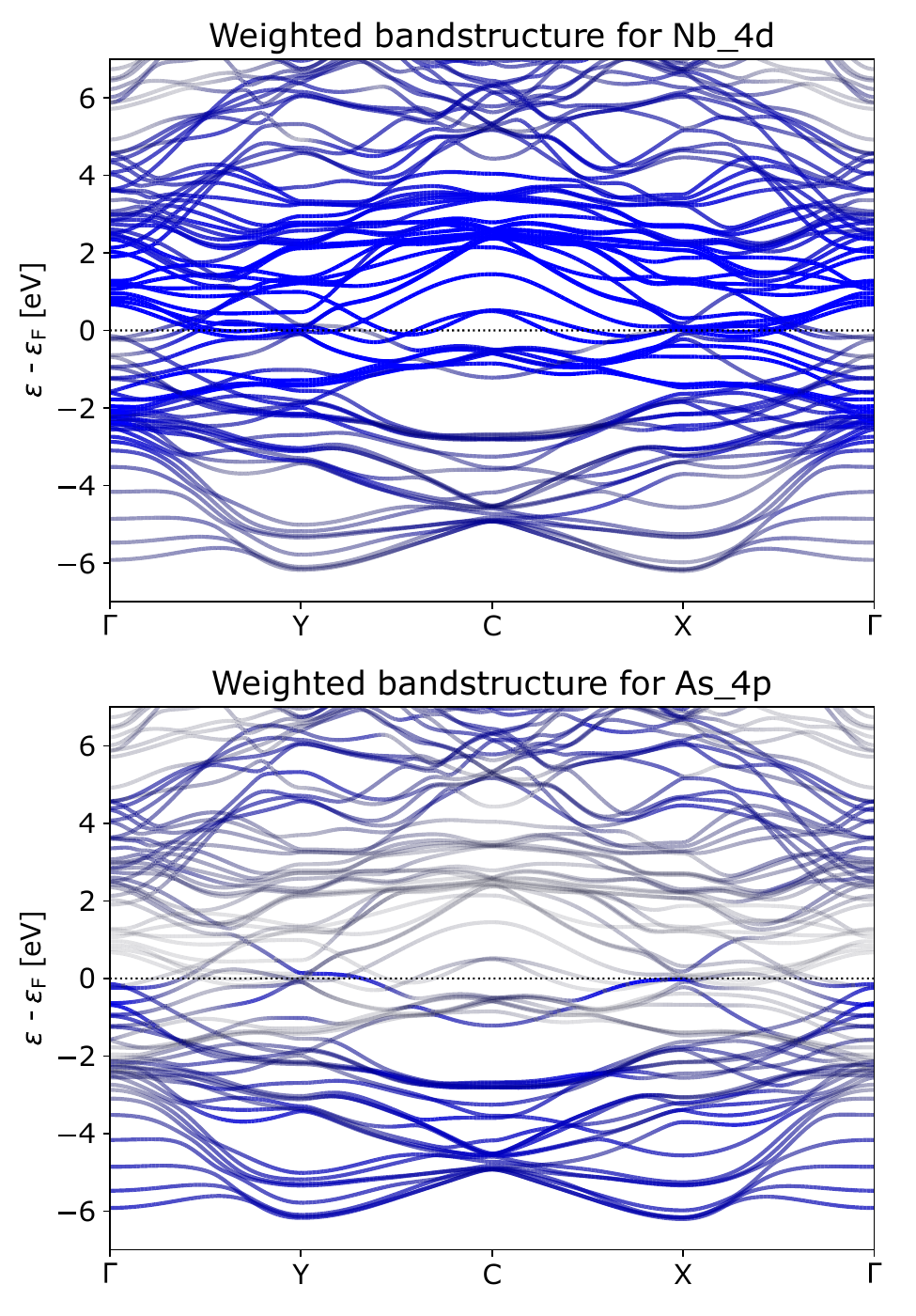}
\caption{Contribution of $d-$orbitals of Nb and $p-$orbitals of As atoms to each band for a 16AL slab}
\label{fig:contrib_bands} 
\end{figure}

\newpage
\begin{figure}[htp!]
\includegraphics[width=\textwidth]{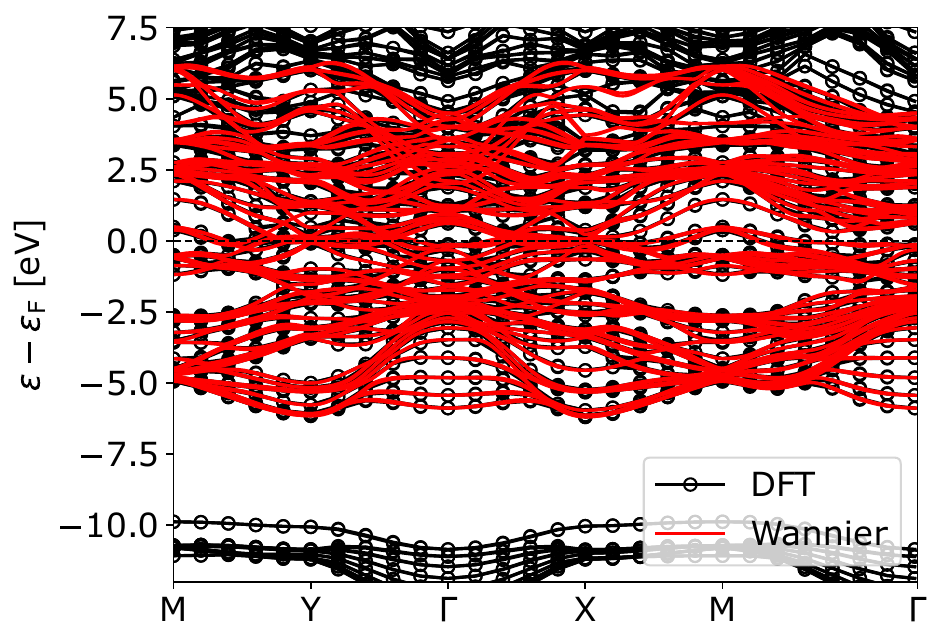}
\caption{Comparison of the DFT and Wannier bandstructures obtained using the basis set of 10 $d-$orbitals per Nb atom and 6 $p-$orbitals per As atom for a 32AL (001)-terminated surface of NbAs slab . }
\label{fig:fermi_surfaces} 
\end{figure}

\newpage
\begin{figure}[htp!]
\includegraphics[width=\textwidth]{schematic view of structure_v3.pdf}
\caption{Schematic view of structure for NEGF calculation for 24AL slab of NbAs with 12-atom defect configuration. The left (L) and right (R) dotted lines denote the semi-infinite contact.}
\label{fig:schematic view} 
\end{figure}

\newpage
\begin{figure*}[htp!]
\centering
\includegraphics[width=\textwidth]{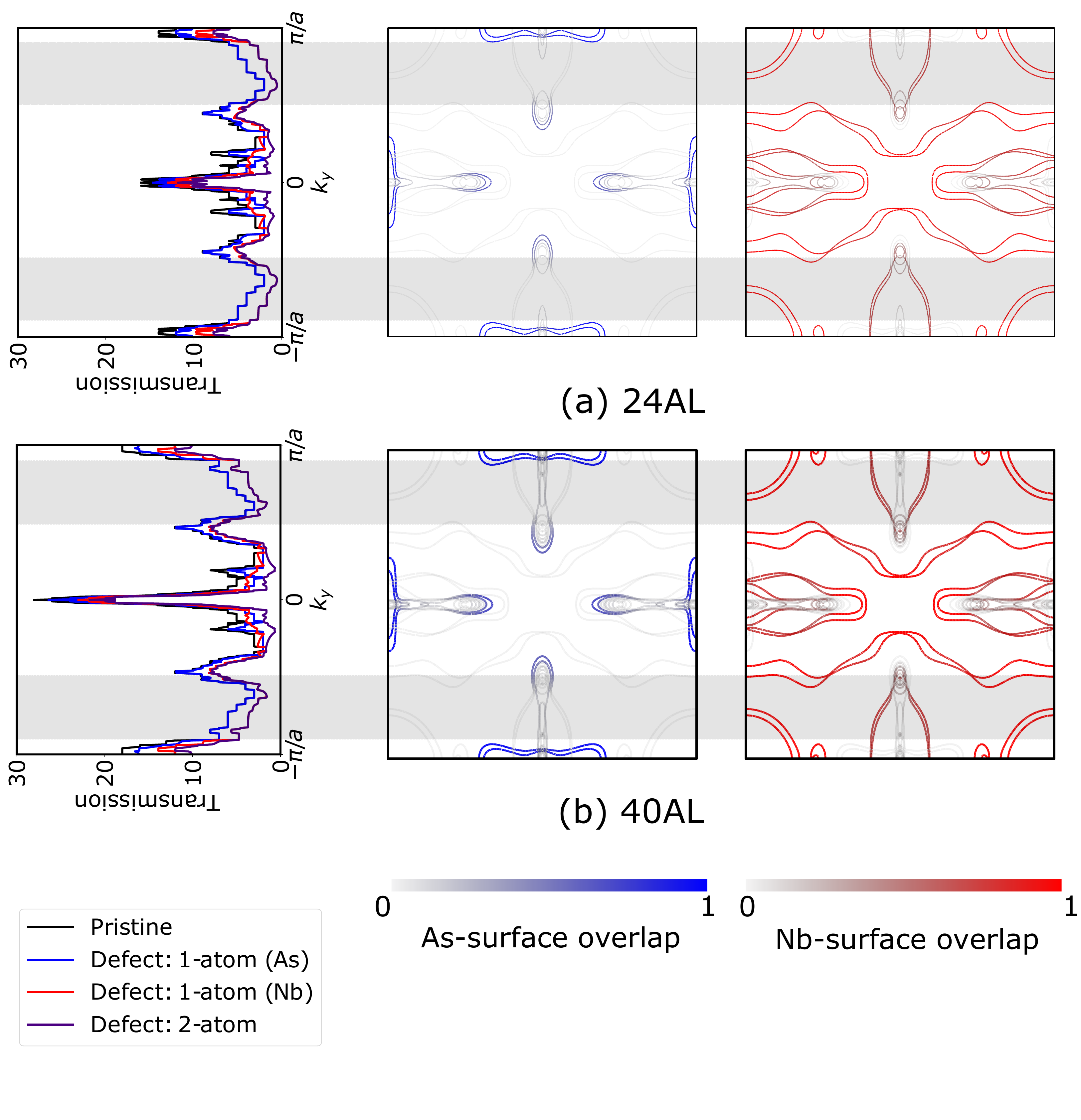}
\caption{Momentum $\mathbf{k}$-resolved transmission for (a) 24AL and (b) 40AL (001) slab of NbAs with different types of defects.}
\label{fig:ballisticG} 
\end{figure*}

\newpage
\begin{table}[htp!]
\begin{center}
\begin{tabular}{ |c|c|c|c|c| } 
\hline
\hline
Defect type & Slope & Intercept & $R^2$  \\
\hline
Pristine	&	0.060	&	3.732	&	0.998	\\
1-atom As	&	0.058	&	3.149	&	0.996	\\
1-atom Nb	&	0.062	&	1.491	&	0.997	\\
2-atom	    &	0.059	&	0.931	&	0.999	\\
6-atom	    &	0.056	&	0.479	&	0.994	\\
12-atom	    &	0.068	&  -0.380	&	0.994	\\
20-atom	    &	0.067	&  -1.112	&	0.994	\\

\hline
\end{tabular}
\caption{\label{tab:parameters}  The slope, intercept and $R^2$ for the linear fits to the transmission vs. thickness data plotted in Figure 5.}
\end{center}
\end{table}

\newpage
\begin{figure*}[htp!]
\centering
\includegraphics[width=\textwidth]{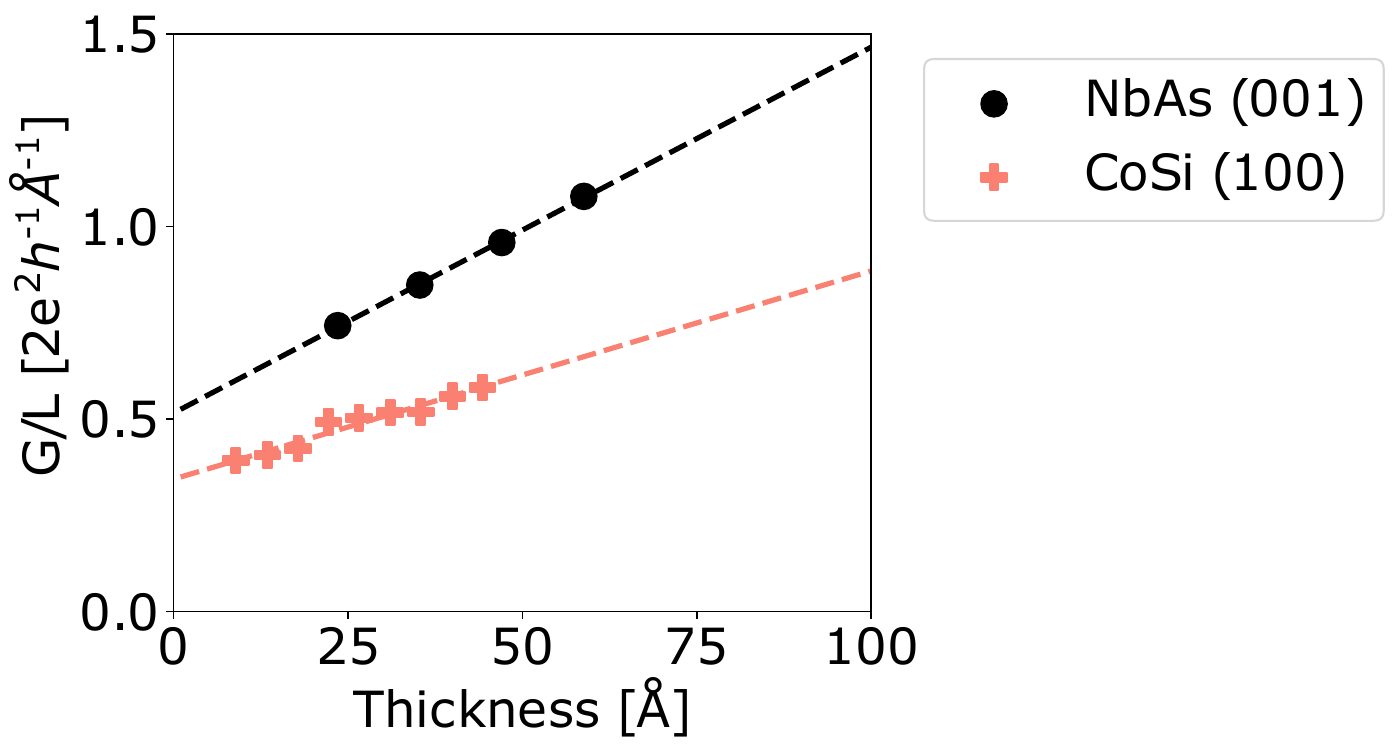}
\caption{Comparison of the total conductance for thin films of NbAs (001) and CoSi (001). }
\label{fig:defect_compare} 
\end{figure*}

\newpage
\begin{figure*}[htp!]
\centering
\includegraphics[width=\textwidth]{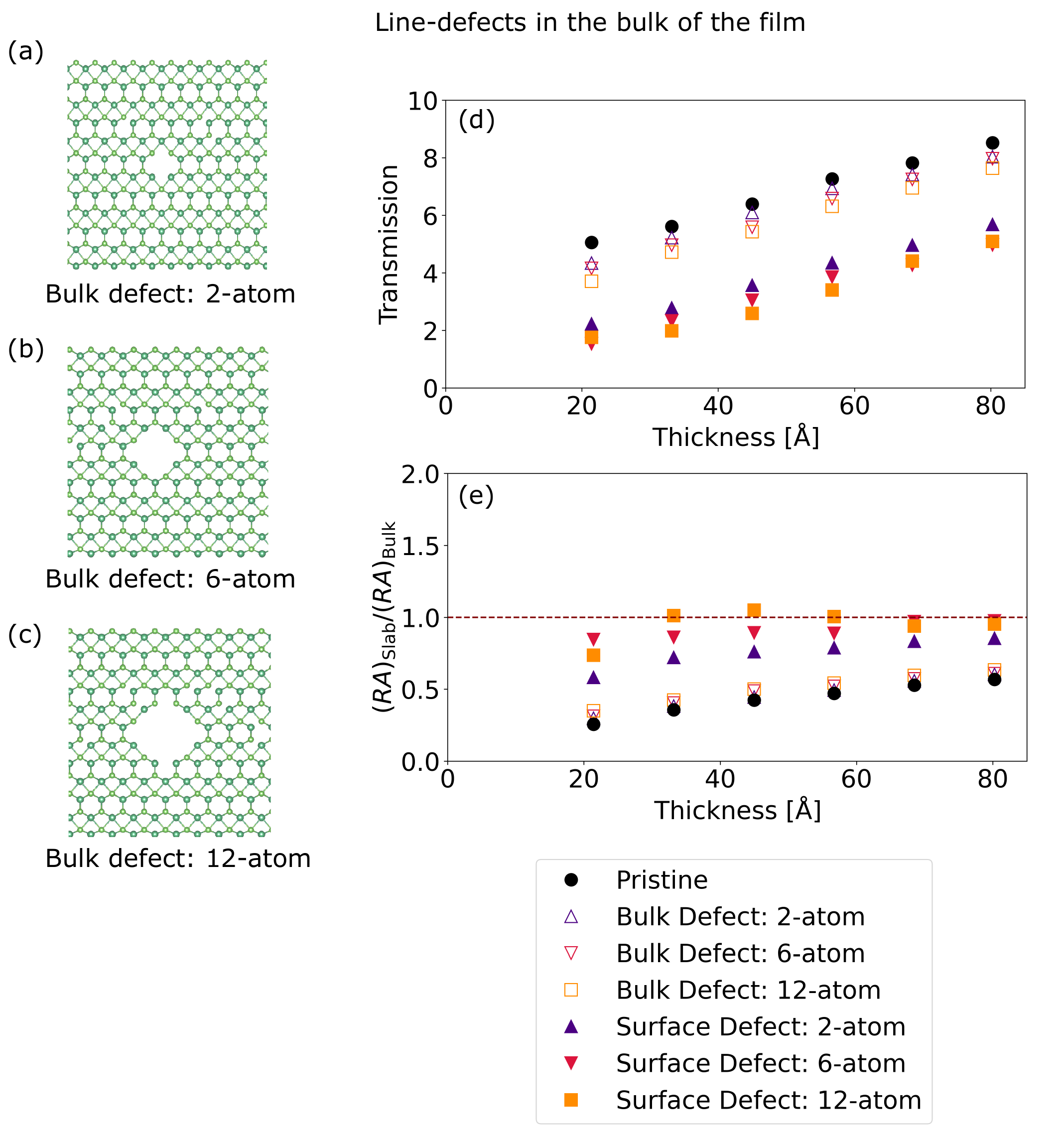}
\caption{Effect of different types of bulk line-defects (a-b) on the transmission (d) and (e) normalized resistance-area $RA$ product of NbAs films.}
\label{fig:defect_compare} 
\end{figure*}

\newpage
\begin{figure*}[htp!]
\centering
\includegraphics[width=0.8\textwidth]{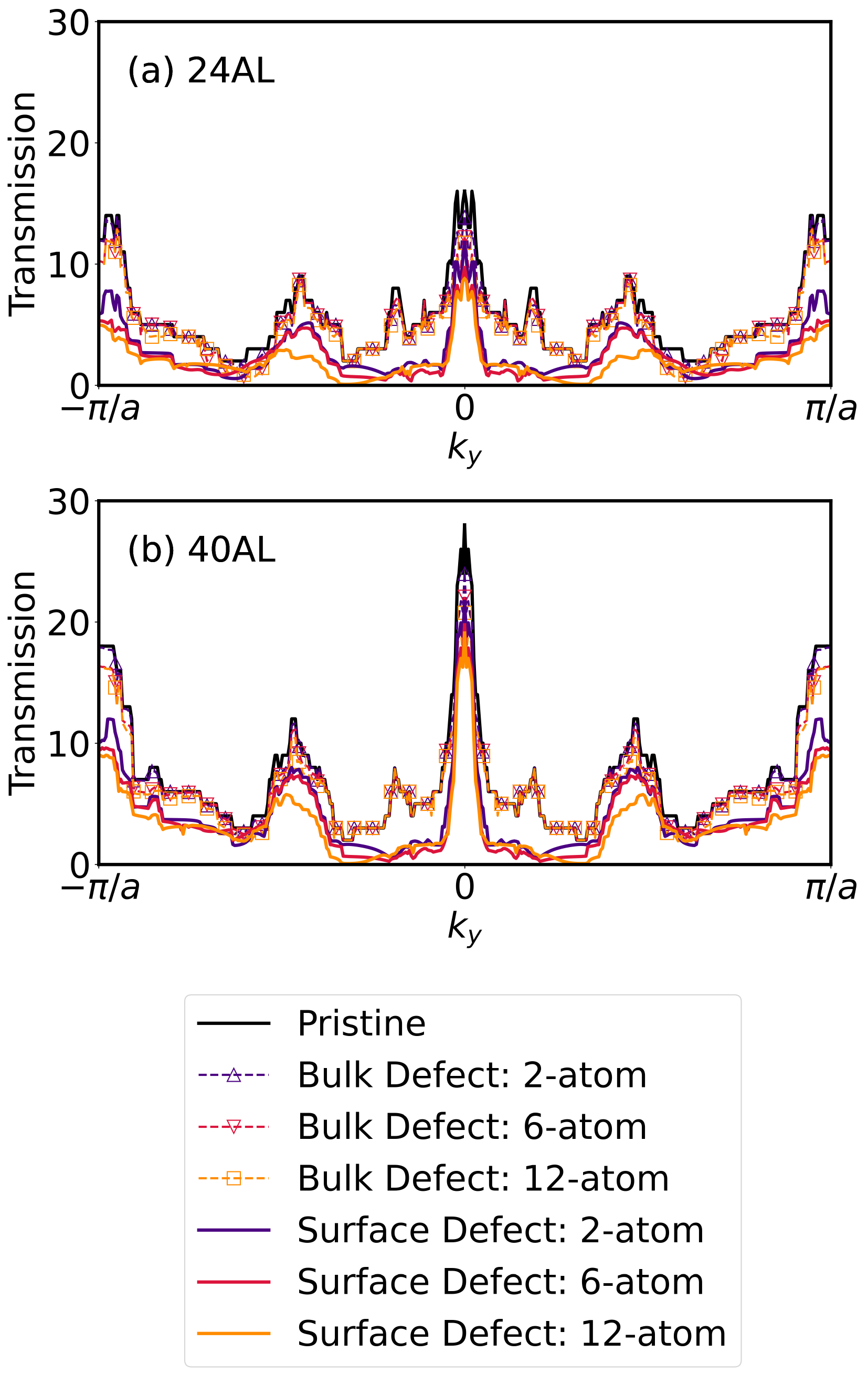}
\caption{Momentum $\mathbf{k}$-resolved transmission for (a) 24AL and (b) 40AL NbAs(001) slabs with bulk and surface line-defects. Bulk line-defects have negligible effect on the transmission across the Brillouin zone showing that transport is dominated by surface states.}
\label{fig:defect_compare} 
\end{figure*}

\newpage
\begin{figure*}[htp!]
\centering
\includegraphics[width=\textwidth]{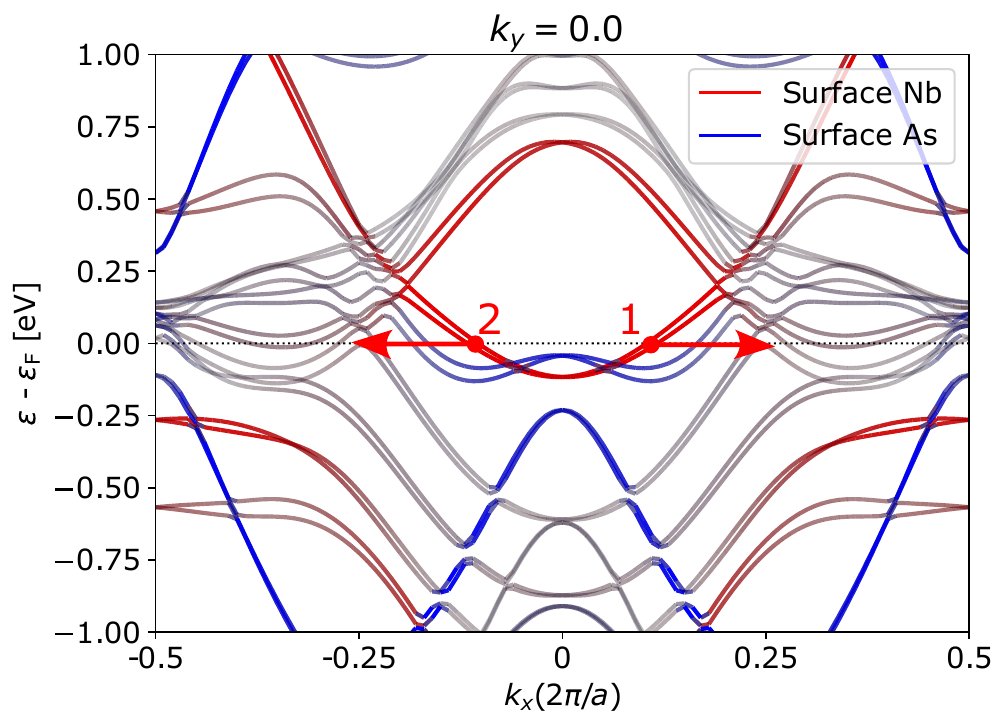}
\caption{DFT bandstructures for 16AL (001) slab of NbAs with colors representing the contributions of the bulk (gray), Nb-terminated (red) and As-terminated (blue) surfaces to the electronic states for different $k_y$ (transverse momentum) in the Brillouin zone. The forward-moving state (1) and backward-moving state (2) are always found to coexist on the same surface. }
\label{fig:states_bandstructure_NbAs} 
\end{figure*}

\newpage
\begin{figure*}[htp!]
\centering
\includegraphics[width=\textwidth]{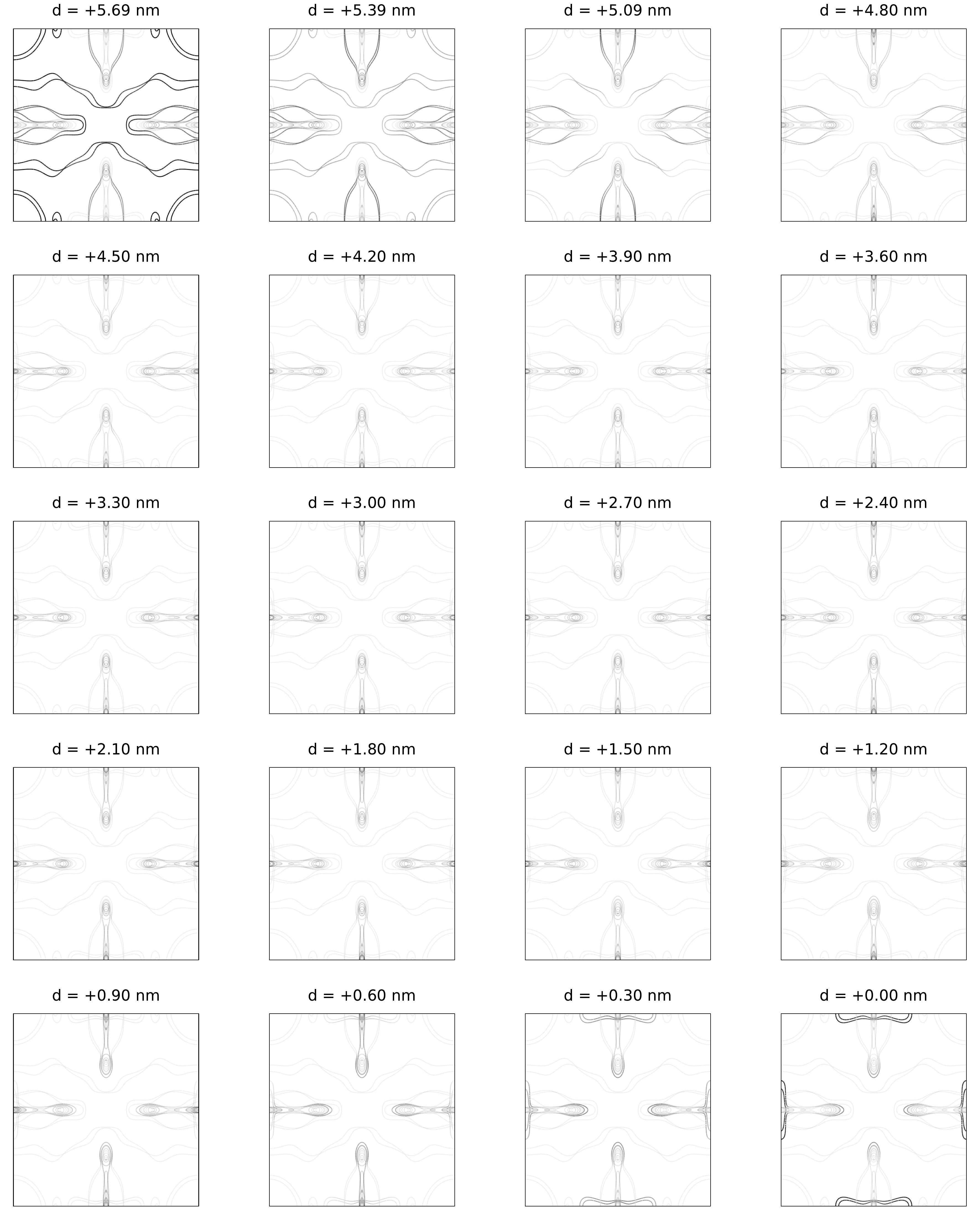}
\caption{Contribution of atomic layers of Nb-As to the Fermi surface of 40AL(001) NbAs slab where $d$ = 0 nm corresponds to the As-terminated
surface and $d$ = 5.69 nm corresponds to the Nb-terminated surface.}
\label{fig:FermiSurface_layers} 
\end{figure*}

\newpage
\begin{figure*}[htp!]
\centering
\includegraphics[width=\textwidth]{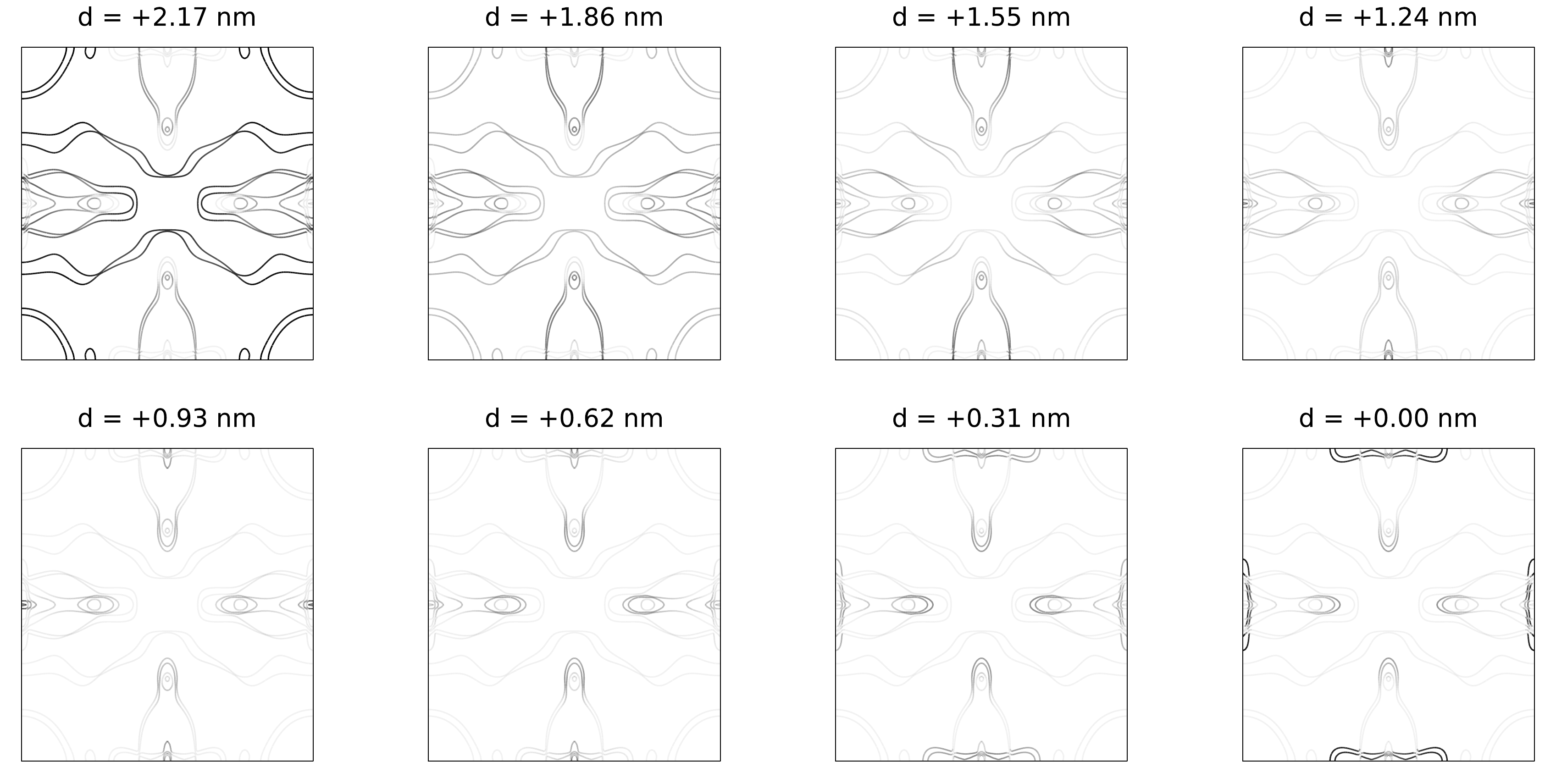}
\caption{Contribution of atomic layers of Nb-As to the Fermi surface of 16AL(001) NbAs slab where $d$ = 0 nm corresponds to the As-terminated
surface and $d$ = 2.17 nm corresponds to the Nb-terminated surface.}
\label{fig:FermiSurface_layers_16AL} 
\end{figure*}

\newpage
\begin{figure*}[htp!]
\centering
\includegraphics[width=\textwidth]{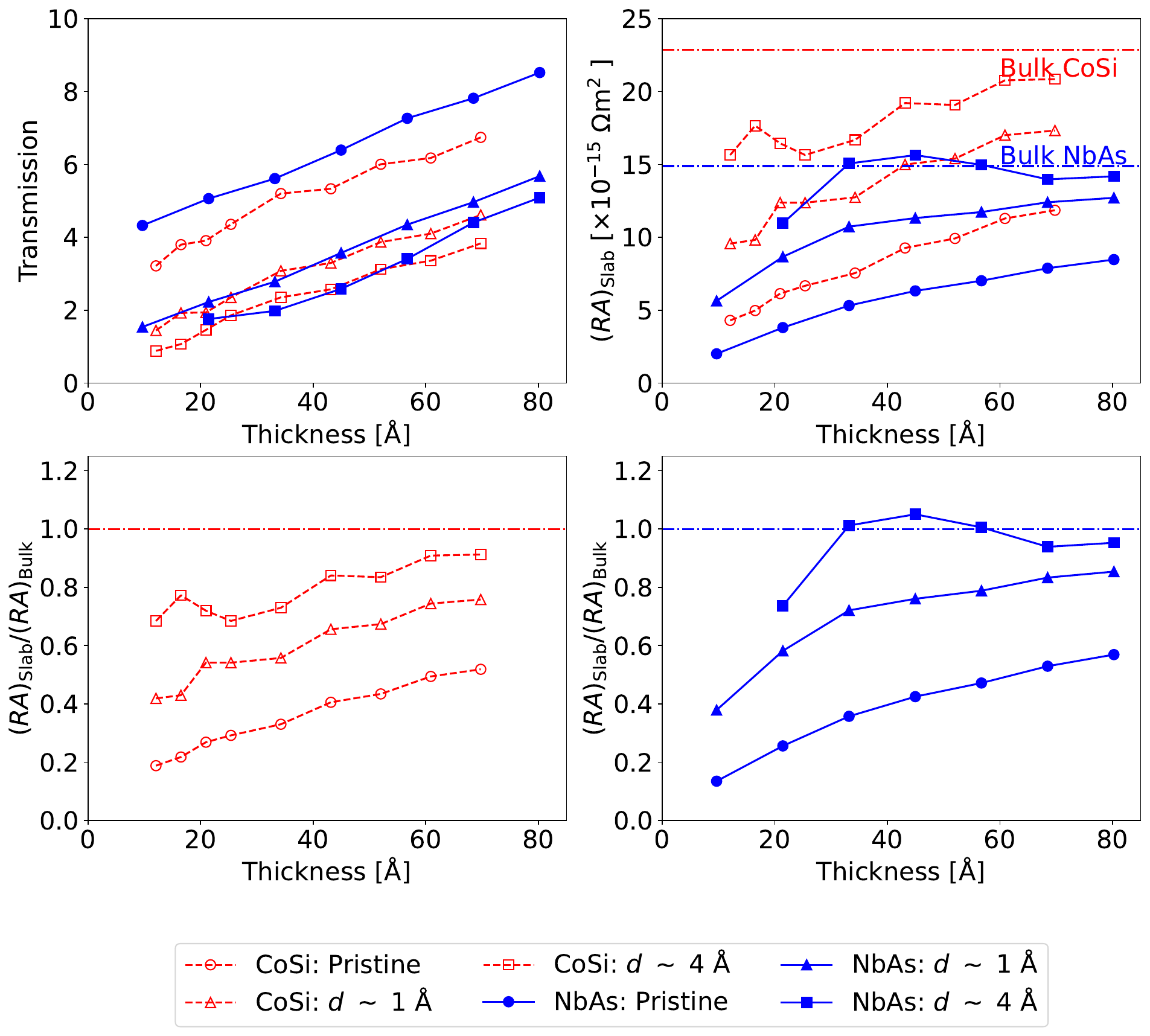}
\caption{(a) Transmission and (b) resistance-area $RA$ product for [100] transport in NbAs(001) and CoSi(001) slabs. The figure shows the impact of surface line-defects of different depths $d$ for each of the two materials.}
\label{fig:FermiSurface_layers_16AL} 
\end{figure*}

 \clearpage 
\bibliography{references}